\documentclass[10pt]{iopart}
\usepackage{bmpsize}
\usepackage{graphicx,color}								
\usepackage{amssymb}
\expandafter\let\csname equation*\endcsname\relax
\expandafter\let\csname endequation*\endcsname\relax
\usepackage{amsmath}
\usepackage{array}
\usepackage{subfigure}
\usepackage{upgreek}
\usepackage[usenames,dvipsnames,svgnames,table]{xcolor}
\newcommand{\ft}[1]{{\color{black} #1}}

\newcommand{\ian}[1]{{\color{black} #1}}

\newcommand{\sconf}{s_{\rm conf}}
\newcommand{\sigmal}{\sigma_{\rm large}}
\newcommand{\sigmas}{\sigma_{\rm small}}
\newcommand{\Nl}{N_{\rm large}}
\newcommand{\Ns}{N_{\rm small}}
\newcommand{\Sideal}{S_{\rm tot}}
\newcommand{\Stot}{S_{\rm tot}} 
\newcommand{\Sexc}{S_{\rm exc}}
\newcommand{\Svib}{S_{\rm vib}}
\newcommand{\Sconf}{S_{\rm conf}}
\newcommand{\etl}{\eta^{\rm loc}}
\newcommand{\cloc}{c^{\rm loc}}
\newcommand{\Slc}{\Delta S_{\rm lc}}
\begin{document}

\title[Experimental Determination of Configurational Entropy in a 2D Liquid under Random Pinning]{Experimental Determination of Configurational Entropy in a Two-Dimensional Liquid under Random Pinning}

\author{Ian Williams$^{1,2,3,4}$, Francesco Turci$^{2,3}$, James E. Hallett$^{2,3}$, Peter Crowther$^{2,3}$, Chiara Cammarota$^5$, Giulio Biroli $^{6,7}$, C. Patrick Royall$^{1,2,3}$}

\address{$^1$School of Chemistry, University of Bristol, Cantock's Close, Bristol, BS8 1TS, UK}
\address{$^2$H.H. Wills Physics Laboratory, Tyndall Avenue, Bristol, BS8 1TL, UK}
\address{$^3$Centre for Nanoscience and Quantum Information, Tyndall Avenue, Bristol, BS8 1FD, UK}
\address{$^4$Department of Chemical Engineering, University of California Santa Barbara, CA 93106-5080, USA}
\address{$^5$King's College London, Department of Mathematics, Strand, London WC2R 2LS, UK}
\address{$^6$Laboratoire de Physique Statistique, \'Ecole Normale Sup\'erieure, CNRS, PSL Research University, Sorbonne Universit\'es, 75005 Paris, France}
\address{$^7$Institut de Physique Th\'eorique, Universit\'e Paris Saclay, CEA, CNRS, F-91191 Gif-sur-Yvette, France}

 \ead{f.turci@bristol.ac.uk}

\begin{abstract}
	A quasi two-dimensional colloidal suspension is studied under the \ian{influence} of immobilisation (pinning) of a \ian{random} fraction of its particles. We introduce a novel experimental method to perform random pinning and, with the support of numerical simulation, we find that increasing the pinning concentration smoothly arrests the system, with a cross-over from a regime \ian{of} high mobility and high entropy to a regime of low mobility and low entropy. At the local level, we study fluctuations in area fraction and concentration of pins and map them to entropic structural signatures and local mobility, obtaining a measure for the local entropic fluctuations of the experimental system.    
	
\end{abstract}


\submitto{\JPCM}

\maketitle
\section{Introduction}

Measuring entropy changes in a liquid is challenging. In molecular liquids, one conventionally proceeds indirectly through calorimetric measurement of specific heat curves. This provides an estimate for bulk entropy changes but yields little information on the structural transformations taking place \cite{ito1999}. The entropy difference between the liquid and the crystalline states of a system,  $\Slc= S_{\rm liq}-S_{\rm crys}$, is at the core of the thermodynamic understanding of the glass transition. When cooling a liquid to its experimental or laboratory glass transition, the difference $\Slc$ becomes approximately constant as the system's structural relaxation time $\tau_{\alpha}$ exceeds the available observation time \cite{debenedetti2001, royall2015physrep}. 

Even more challenging is to disentangle configurational contributions to the entropy (which account for the number of disordered \textit{inherent} states \ian{accessible to} the liquid) from the vibrational ones. \ian{On approaching} the experimental glass transition temperature, $\Slc$ is dominated by the difference in configurational entropy, as quantitatively demonstrated in several computer simulations \cite{sciortino1999,stillinger2002energy,sciortino2005,turci2017nonequilibrium}. In the thermodynamic interpretation of the glass transition of Adam-Gibbs (inherited by the Random First Order Transition (RFOT) framework \cite{berthier2011}), the entropy change $\Slc$ \ian{vanishes} at a finite temperature, $T_K$, with diverging structural relaxation times. However, such a scenario is \ian{experimentally} inaccessible due to \ian{an unavoidable} glass transition, which leads the \ian{supercooled} liquid off-equilibrium \ian{into} an ageing glassy state. Hence, within the RFOT framework, \ian{the introduction of an additional parameter to control entropy reduction} has been proposed \cite{cammarota2012}. Numerical studies \cite{biroli2008,berthier2012,cammarota2013general,fullerton2014investigating,ozawa2015} of supercooled liquids with a finite fraction, $c$, of immobile \textit{pinned} particles have shown that it is possible to reduce the bulk configurational entropy of the liquid at moderately low temperatures and to reveal -- in three spatial dimensions -- a first-order transition with near-critical fluctuations in the degree of similarity or \textit{overlap} between different configurations, the central order parameter of the RFOT framework. \ian{In two dimensions,} this \ian{transition} is expected to mutate \ian{into} a cross-over \cite{cammarota2013general} between a free regime at low concentrations \ft{of pinned particles} and a frustrated, immobilised regime at high concentrations. However, while the introduction of a fraction of pinned particles as a control parameter is trivial in computer simulations, \ian{it} is much more difficult \ian{to realise in} experiments.

Colloidal experiments are in this sense most promising: \ian{individual} particles can be \ian{observed and their trajectories tracked in time}, allowing direct comparison with molecular simulations. Recent \ian{colloidal experiments have shown} it \ian{is} possible to emulate the effect of pinned particles \ian{using} holographic optical tweezers \ian{to} apply a local confining potential to individual particles \cite{gokhale2014,gokhale2016,gokhale2016jsm}. \ian{This} technique is very powerful as it allows \ian{precise} control \ian{of pinned particle locations} and their \ian{spatial} distribution. \ian{However, it} presents some drawbacks, \ian{key amongst them being limited to only} tens of \ian{simultaneously immobilised} particles, hindering the statistics. \ian{Further complications arise due to interference between multiple optical traps resulting in spatial intensity variations and weak, unwanted ``ghost traps'', creating uncertainty in the optical energy landscape applied to the system \cite{Bianchi2010,Bowman2011}.}

Here we \ian{introduce an alternative} approach. We investigate the effect of pinning in a quasi-two dimensional binary mixture of hard colloidal particles \ian{sedimented against a glass substrate \cite{Thorneywork2014,Gray2015}}. Immobilisation occurs at random due to \ian{particles overcoming the repulsive electrostatic barrier at the substrate and entering an attractive van der Waals minimum resulting in their adsorption and pinning to the glass surface}, see Fig. \ref{figScheme} \ian{(c)}. \ian{Consequently, one observes} a progressive, irreversible increase \ian{in} the concentration of pinned particles. \ian{The spatial distribution of pinning sites is} not externally controlled and is, in this sense, random. Through optical microscopy, we \ian{obtain} particle \ian{trajectories} and measure the effect of progressive immobilisation on the local structure. \ian{Experimental studies are augmented with} numerical simulations, mapping local area \ian{fraction} and pinned particle fluctuations to a local order parameter derived from the bulk configurational entropy and measuring local fluctuations in the two-body excess entropy.  While, at the bulk level, configurational entropy reduction and \ian{increasing} relaxation times evolve similarly with increasing \ian{pinned} fraction, at the local level, low entropy regions appear to freeze-in local area fraction and pinned-particle fluctuations without strong correlations with the local mobility. 
	
The article is organised as follows: Section \ref{secExp} \ian{describes} the experimental approach. Section \ref{secSim} reports the numerical model of the system, with \ian{further} details on the entropy calculations in the Appendix. In Section \ref{secEquilibrium} we discuss the equilibrium phase diagram of the system, as obtained from simulations, and relate it to the experimental observations. In Section \ref{secNoneq} we describe the effect of a slowly increasing fraction of pinned particles on the mobility and fluctuations in the local two-body excess entropy. \ian{Finally we conclude and summarise our findings and} discuss their implications for future work.
	
\section{Experimental methods}
\label{secExp}
\subsection{Protocol}
\begin{figure}[t]
\centering
	\includegraphics[scale=1]{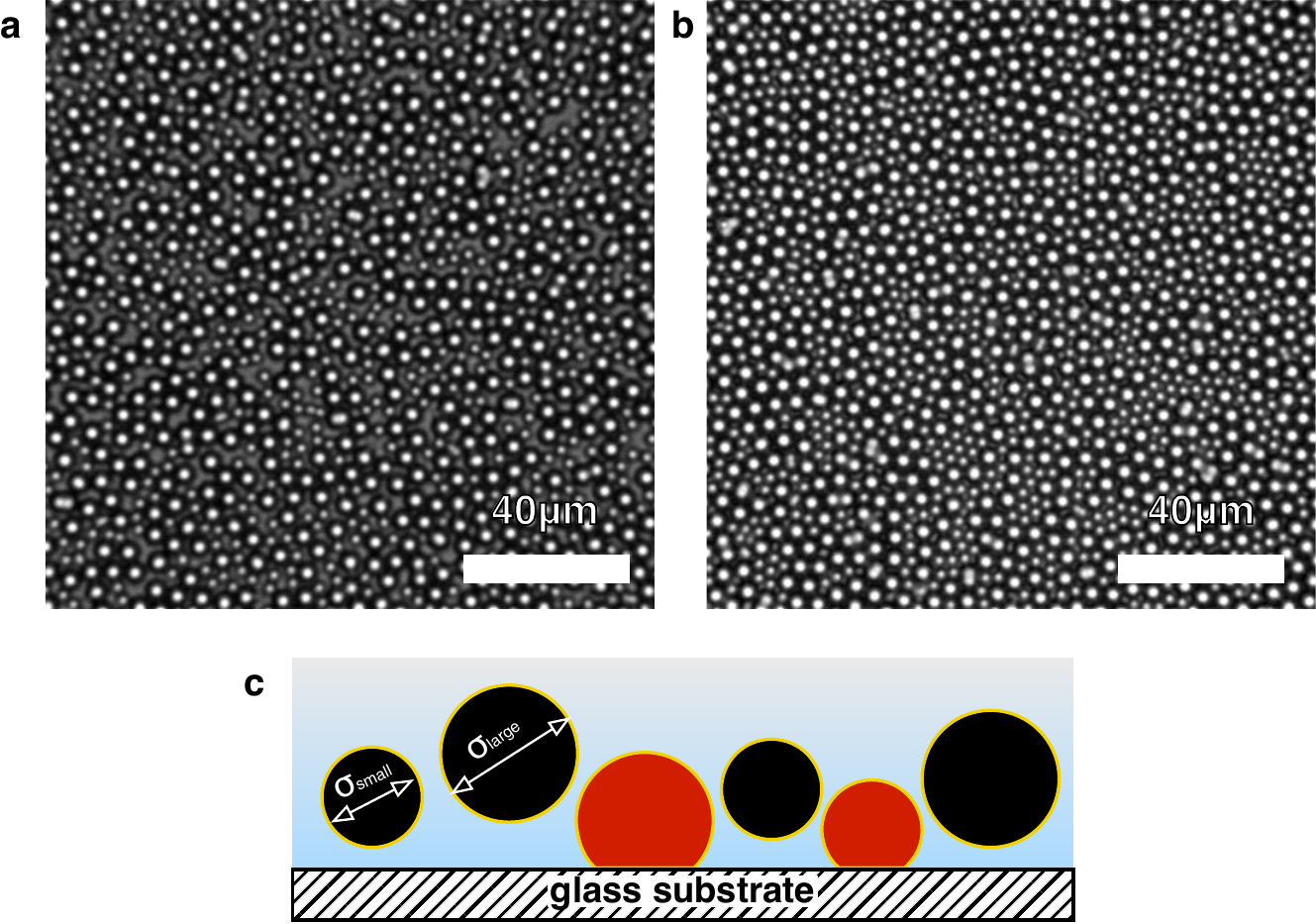}
	\caption{ Experimental system: (a) and (b) are micrographs showing the experimental system at overall area fractions $\eta = 0.65$  and $\eta = 0.88$ respectively. (c) Schematic representation of the experimental setup. Binary colloidal silica particles of two characteristic sizes $\sigmas = 3 \; \mathrm{\upmu m}, \sigmal = 5 \; \mathrm{\upmu m}$ form a quasi-2d \ian{layer} adjacent to a glass coverslip. They occasionally \ian{overcome the repulsive electrostatic barrier} with the glass substrate and are immobilised \ian{due to van der Waals attraction} (red particles).  }
	\label{figScheme}
\end{figure}

Optical microscopy \ian{is} performed using a Leica SP5 microscope in brightfield mode. The experimental system is a 50:50 number ratio mixture of $5.0\; \upmu \mathrm{m}$ and $3.0 \; \upmu \mathrm{m}$ diameter silica spheres suspended in deionised water. \ft{We consider a binary mixture in order to minimise the emergence of hexatic/hexagonal order and choose the corresponding size ratios in order to limit fractionation of the two species. This allows us to study disordered systems even at very large area fractions.} Due to the density mismatch with the solvent, particles sediment to form a monolayer adjacent to a glass coverslip \cite{Gray2015,Tamborini2015}. The gravitational lengths of $5 \; \mathrm{\upmu m}$ and $3 \; \mathrm{\upmu m}$ silica spheres in water are $3.9 \; \mathrm{nm}$ and $18 \; \mathrm{nm}$ respectively, and so the system can be considered quasi-two-dimensional. After sedimentation, area fractions in the range $0.6 < \eta < 0.88$ are obtained \ian{as shown in the micrographs in Fig.~\ref{figScheme}(a) \& (b)}. \ian{The suspension is stabilised against aggregation by an electrostatic interparticle repulsion, however, this is sufficiently short-ranged that} the addition of salt has a negligible effect on the interparticle potential \cite{Cui2002}, and so we treat our sample as binary quasi-hard-discs. The monodisperse hard disc system exhibits \ian{a first order transition from an isotropic liquid to a} hexatic structure \ian{at} $\eta = 0.7$ and \ian{a subsequent continuous transition to the} hexagonal crystal \ian{at} $\eta = 0.72$ \cite{bernard2011}. However, long-range and quasi-long-range orientational and translational ordering \ian{is inhibited in our binary system}, maintaining a liquid-like structure up to the highest area fractions considered. Experiments are performed on a \ian{precisely} levelled optical table to prevent \ian{systematic lateral drift due to} sedimentation parallel to the imaging plane. We note that, \ian{following sedimentation, the equatorial planes of large and small particles are at different heights and so} the \ian{true interparticle} interactions \ian{are} non-additive \cite{assoud2010}. \ian{The distance of closest approach between a large and small particle is given by $\sqrt{\sigmal + \sigmas}$, which for our system is reduced by $3.2\%$ compared to the in-plane approach distance $(\sigmas + \sigmal)/2$. Thus, non-additivity is a small effect and is neglected} in the following modelling and discussion.

%

Over time, \ian{particles can overcome the electrostatic repulsion from the substrate and} van der Waals \ian{attractions cause irreversible adsorption. Thus, the fraction of immobilised particles increases with time.} Typically, this is suppressed by treating the glass surface with \emph{e.g.} a silane. \ian{Here, however, we exploit} this effect to obtain a \ian{quasi-}static subpopulation of pinned particles. This is illustrated in Fig.~\ref{figScheme}(c). \ian{A range of pinning densities is obtained by varying the waiting period} between sample preparation and observation. Longer waiting times yield greater \ian{concentrations} of pinned particles. The rate of pinning on untreated glass coverslips is sufficiently fast that the pinned density cannot be considered quasi-static over the timescales of image acquisition. Therefore, we slow the rate of pinning by treating the substrate with \ian{low concentration solutions of} Gelest Glassclad 18. \ian{With suitable treatment,} we find that pinning density is approximately constant over periods of 1 hour, but \ian{not entirely suppressed, growing} over \ian{a timescale} of days. \ft{This allows us to tune the growth rate of the pinning concentration and obtain both quasi-stationary growth curves (representative of equilibrium) and nonstationary curves, in full nonequilibrium conditions, see Fig. \ref{figPinIdent}(a)}.


\subsection{Identifying pinned particles}

\begin{figure}[t]
\centering
	\includegraphics[width=\textwidth]{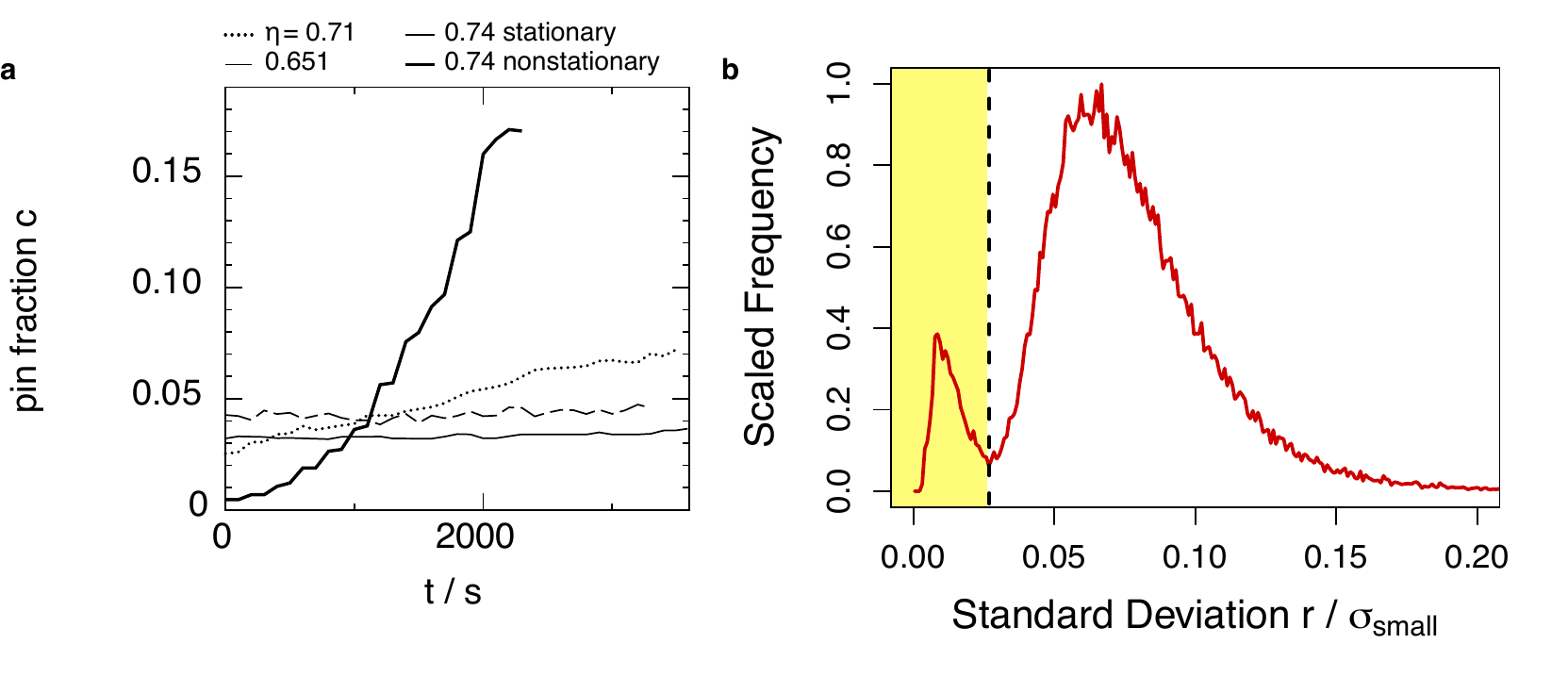}
	\caption{(a) Time evolution of the detected \ian{fraction} of pinned particles in different experiments. Depending on the \ian{glass surface treatment} the pinning concentration can be held stationary or can increase dramatically \ian{over the duration of an experiment}. (b) Histogram showing standard deviation in particle positions over time interval $\Delta t = 60 \; \mathrm{s}$ in an experiment at area fraction $\eta = 0.71$. Two populations corresponding to pinned (small standard deviation, shaded region) and unpinned particles (large standard deviation) are observed, and thus pinned particles are identified.}
	\label{figPinIdent}
\end{figure}

Particle trajectories are obtained using routines based on those of Crocker and Grier, implemented in the R programming language \cite{crocker1995,Gray2015}. Large and small particles are distinguished based on the integrated brightness of their images. Subsequently, trajectories are used to identify the pinned subpopulation. When a particle attaches to the substrate it stops moving. Thus, identifying a particle as pinned requires an inherently dynamic criterion. Since there is \ian{unavoidable} error in locating the centre of a particle image due to pixel noise, even pinned particles show some small but nonzero displacement between frames. Thus, in order to identify pinned particles, each trajectory in a given experiment is split into subtrajectories of length $\Delta t$. Within each subtrajectory the standard deviation in particle position is calculated. The histogram of these standard deviations for all particles in a given experiment reveals two peaks, as shown in Fig.~\ref{figPinIdent}(b) --- pinned particles have a much smaller standard deviation in position than their unpinned counterparts. A cut-off is applied at $0.027 \sigmas$ and all particles exhibiting a standard deviation smaller than this value are considered pinned particles. The value of $\Delta t$ is chosen independently for each experiment, as at higher area fractions, particles motion is increasingly hindered by a particle's neighbours. In practise, $\Delta t$ is chosen so that a clear distinction between the peaks corresponding to pinned and unpinned particles is evident in the histogram. For the lowest density experiments $\Delta t = 60 \; \mathrm{s}$ while at the highest densities, $\Delta t = 1000 \; \mathrm{s}$.


\subsection{Advantages and disadvantages of the adsorption pinning technique}

\begin{figure}[hbt]
	\centering
	\includegraphics{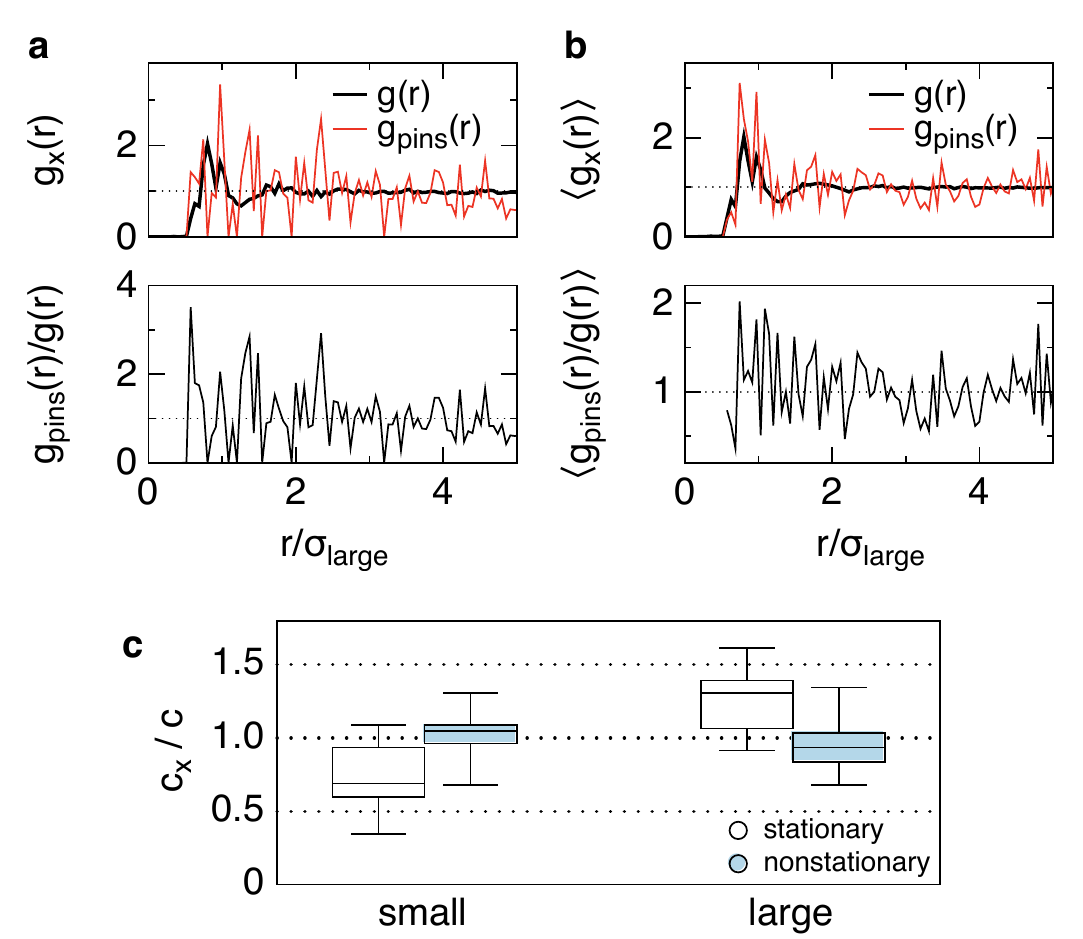}
	\caption{(a,b) Pair correlation functions for the pinned particles $g_{pins}(r)$ and for all of the particles $g(r)$: (a) at $\eta = 0.711$ and $c=0.05$; (b) averaging over an ensemble of configurations with $0.65<\eta< 0.738$ and $0.03<c<0.08$. (c) Box-plot of the ratios of the fraction $c_x$ of pinned particles of type $x$ and the overall fraction $c$ for large and small particles in stationary (white) and nonstationary experiments (shaded). }
	\label{figgsPins}
\end{figure}

\ft{Our experimental methodology allows us to obtain two-dimensional disordered packings of spheres, with moderate (below $c=0.10$ ) to large (beyond $c=0.10$ and up to $c=0.90$) fractions of pins. When working in the moderate pinning regime, it is relatively easy to maintain a stationary pinning concentration (see Fig.~\ref{figPinIdent}(a)). In this regime, the obtained pinned particles are distributed approximately uniformly on the plane. We show this through the analysis of pair correlation functions, Fig.\ref{figgsPins}. We compute the instantaneous pair distribution function}
\begin{equation}
	g_x(r) =\frac{1}{\rho_x} \sum_{i\neq j}\delta(|\vec{r}_i-\vec{r}_j| -r)
\end{equation}
\ft{where $\rho_x$ is the density of particles in the set $x$ and $i,j$ are indices running over all the considered particles. We focus on the total radial distribution functions $g(r):=g_x|_{x:\{ \textrm{all the particles} \} }$ and the pinned radial distribution function $g_{\rm pins}(r):=g_x|_{x:\{ \textrm{only the pinned particles} \} }$. For the former we can perform a time average while for the latter this is evidently not possible, as pinned particles are by definition arrested. In Fig.~\ref{figgsPins}(a) we see how the two pair distribution functions  compare at a given value of area fraction and concentration of pins.}

\ft{Hence, to average the fluctuations, we observe that at sufficiently high area fractions the pair correlation functions varies slowly and we can average the ratio $g_{\rm pins}(r)/g(r)$ in an interval of area fractions and concentration of pinned particles. In Fig.~\ref{figgsPins} (b) we plot these averages and notice that the radial distribution of pinned particles follows closely the behaviour of the total pair distribution function. Thus, in this regime, we can consider the pinned subpopulation to be random, with structural features that are indistinguishable from the overall particle population.} 


\ft{Furthermore, if pinning is truly random, one expects the large-small ratio within the pinned subpopulation to reflect that of the sample as a whole, implying $N_{\rm small}^{\rm pinned}/ N_{\rm small}=N_{\rm large}^{\rm pinned}/N_{\rm large}$.
We compute the ratio $N_x^{\rm pinned}/cN_x=c_x/c$ for $x:\{ \textrm{large, small} \}$, where $N_x$ is the number of particles of type $x$ and $c_x$ indicates the fraction of particles of type $x$ that is pinned. We scale this by $c$, the average total fraction of pinned particles. A value close to unity corresponds to uniform pinning. As shown in Fig.~\ref{figgsPins}(c), stationary conditions display a statistically relevant prevalence of pinned large particles compared to small particles. When the pinning is nonstationary, however, small and large particles are more evenly pinned.} \ft{We interpret this effect as a consequence of the fact that larger particles sediment faster than smaller particles and they are more prone to be adsorbed due to their shorter gravitational length.} 

\ft{Our protocol allows the easy preparation and observation of large, two-dimensional disordered systems with a finite fraction of immobilised particles. However, it also presents some important drawbacks. The method as described here is limited to two-dimensional systems.} \ian{Numerical studies \cite{biroli2008,berthier2012,cammarota2013general,fullerton2014investigating,ozawa2015} predict that the three dimensional first order pinning glass transition is reduced to a weak crossover in two dimensions, meaning our technique is limited in its ability to probe theoretical predictions. Secondly, while the extent of glass surface treatment facilitates some control over the rate of particle pinning, this control is limited, and targeting a specific rate \emph{a priori} remains an experimental challenge. Finally, dense, supercooled suspensions are by their nature very slow, and so equilibration on the experimental timescale becomes impossible at high density. Keeping experiments short enough that the pin fraction is quasi-static but allowing the unpinned particles sufficient time to explore their free energy landscape and relax is a significant challenge at high densities.} 

\section{Numerical simulations}
\label{secSim}
\subsection{Molecular dynamics simulations}
We model the interactions between hard spherical colloids \ian{using} the pseudo hard sphere potential, a piecewise continuous truncated and shifted variant of the Mie potential \cite{jover2012p,espinosa2013}:

\begin{equation}
\label{eq:potential}
 	\beta V(r)=\begin{cases}
 	m \left(\frac{m}{n} \right)^{n}\beta\epsilon  \left[\left(\frac{\sigma_{ij}}{r}\right)^{m} -\left(\frac{\sigma_{ij}}{r}\right)^{n}\right]+\beta\epsilon,  &  r<\frac{m}{n}\sigma_{ij} \\
       0, &\text{otherwise}
        \end{cases}
\end{equation}
with $\sigma_{ij}=(\sigma_i+\sigma_j)/2$ between particles $i,j$, with diameters $\sigmal/\sigmas=5/3$ and $\beta\epsilon=\epsilon/k_BT=2/3$ \ft{with $k_B$ set to unity} and exponents $m=50,n=49$, for which the system behaves as a bidisperse mixture of hard spheres in three dimensions and hard disks in two dimensions. We model an equimolar mixture with $\Nl=\Ns$ with total particle number $N=\Nl+\Ns$, ranging from $6\cdot 10^3$ to $6\cdot 10^4$ . The area fraction is defined as $\eta=\pi\rho(\sigmal^2+\sigmas^2)/8$, with number density $\rho=N/L^2$, where $L$ is the  lateral size of a two-dimensional square simulation box. \ian{Periodic boundary conditions are employed throughout in order to simulate bulk behaviour.} \ft{Our model neglects hydrodynamic interactions, which, although important in the formation of colloidal bonds in other physical processes such as gelation, nucleation and yielding \cite{ness2017effect,varga2016hydrodynamic}, play no role in the dense, hard system considered here.} 

\ian{This} model is convenient as it allows us to run Molecular Dynamics simulations in the isochoric-isothermal ensemble (NVT), applying a velocity-Verlet integrator of timestep $dt=0.0015\sqrt{\sigmas^2/\epsilon m_{\rm small}}$ coupled to a Nos\'{e}-Hoover thermostat of damping time $t_{\rm damp}=0.1\sqrt{\sigmas^2/\epsilon m_{\rm small}}$ using the LAMMPS molecular dynamics package \cite{plimpton1995}. Molecular dynamics allows us to easily obtain accurate estimates of the pressure from the virial expression, which are useful when computing the configurational entropy via thermodynamic integration.  

Pinned particles are randomly selected on the plane, regardless of their size or relative position, \ian{mimicking} the experiment. When a non-zero fraction of pinned particles is considered, the thermostat is only applied to the mobile particles, while velocities and accelerations are zeroed for all the pinned particles.

\subsection{Entropy calculations and representations}

Following previous works \cite{ozawa2015,Angelani2005,angelani2007}, we compute the configurational entropy via thermodynamic integration, effectively considering the particles as hard disks. To this \ian{end}, we employ \ft{an approach \cite{Angelani2005} based on the} Frenkel-Ladd method \cite{frenkel} to obtain the vibrational part of the entropy $S_{\rm vib}$ from the mean square displacement with respect to a reference configuration. The configurational entropy \ian{is calculated by subtracting} this from the total entropy, $S_{\rm tot}$, \ian{which is} obtained from the equation of state for the pressure as a function of area fraction for several pinning fractions $P(\eta;c)$. More details can be found in the Appendix.

From the equilibrium estimates of the configurational entropy we define a ``local entropy''. For every \ian{configuration,} we compute the corresponding radical Voronoi tessellation (which takes into account the polydispersity of the system) and from the tessellation we \ian{define} the local area fraction as 
\begin{equation}
	\etl_i=\frac{a_i+\sum_{j\in {\rm neighbors}}^{N_i} a_j}{v_i+\sum_{j\in {\rm neighbors}}^{N_i}v_j},
\end{equation}
where the sum runs over the $N_i$ Voronoi neigbors of particle $i$, \ian{$a_x$ represents the area of particle $x$ and $v_x$ is the area of its Voronoi cell.} Similarly we define a local fraction of pinned particles $\cloc_i$ from the local fraction of particles (including particle $i$) that are pinned. We then associate a local configurational entropy through interpolation, \begin{equation}
s^{\rm loc}_i= \sconf(\etl,\cloc).
\label{eq:sloc}
 \end{equation}
By this means, we perform a non-linear mapping of the local area fraction and local fraction of pins to \ian{obtain} a 2d representation of the regions with higher/lower configurational entropy.

An alternative measure of the local entropy is provided by the local two-body fluctuations of the radial distribution function, contributing to the so-called two-body excess entropy, $\tilde{s}$.  It has recently been shown \cite{piaggi2017} that, for monodisperse crystalline systems, this provides an interesting local fingerprint of emerging order. For our binary mixture we follow \cite{piaggi2017} and define a \ft{smooth} pair distribution function for every particle $i$:
\begin{equation}
g^i_{\alpha\beta}(r)=\frac{1}{2\pi r}\sum_j \frac{1}{\sqrt(2\pi \delta^2)}e^{-(r-r_{ij})^2/(2\delta^2)},
\end{equation}
between species $\alpha,\beta$ with $\delta=0.12\sigmas$, where the sum runs over a shell of neighbours \ian{within a} fixed radius $r_{m}=5\sigmas$. We employ this definition to compute, for every particle, a local entropic signature,
\begin{equation}
s_2^i(\rho)=-k_B\frac{\rho}{2}\sum_{\alpha\beta}x_{\alpha}x_{\beta}\int \left[g^i_{\alpha\beta}(r)\ln g^i_{\alpha\beta}-g^i_{\alpha\beta}+1\right]r dr,
\end{equation}
which we average locally over the Voronoi neighbours of each particle to obtain
\begin{equation}
\tilde{s}^i=\frac{s_2^{i}+\sum_{j \in{\rm neighbours}} s_2^j}{1+N_i},
\label{eq:stwobody}
\end{equation}
where $N_i$ are the Voronoi neighbours of particle $i$.

\begin{figure}[t]
	\centering
	\includegraphics{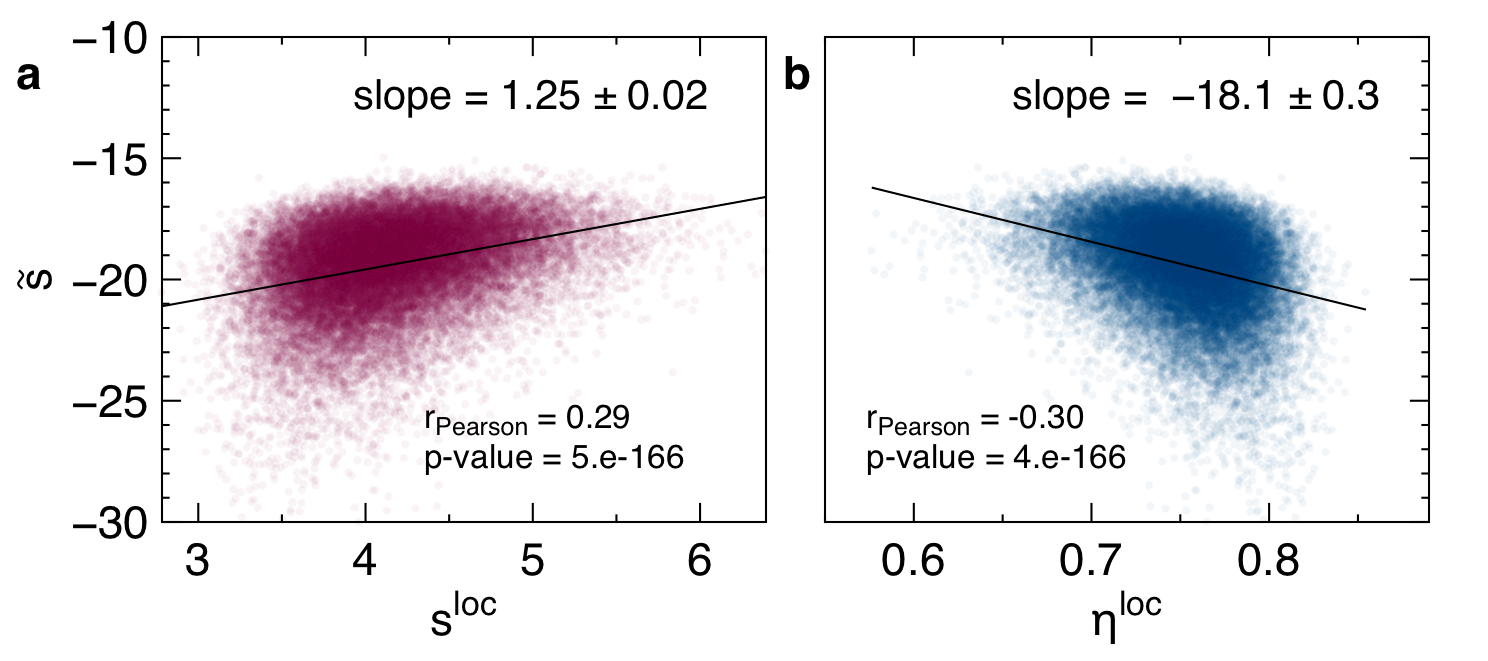}
	\caption{Correlations between the local two-body entropy and (a) the local entropy (b) the local area fraction from radical Voronoi tessellation at total area fraction $\eta=0.75$ from numerical simulations.}
	\label{figScorrelation}
\end{figure}

We show in Fig.~\ref{figScorrelation} that such a measure of local two-body entropy is positively correlated with the local entropy obtained from interpolation, $s^{\rm loc}$ (Eq.~\ref{eq:sloc}), and negatively correlated with the area fraction, with very small p-values indicating that the correlation, albeit moderate (Pearson coefficients of 0.29 and -0.30 respectively), is rather robust \ft{(the error on the slope is below 2\%)}. 

\ft{Note that even if $s^{\rm loc}$ and $\tilde{s}$ are correlated, they are two radically different quantities: the former is obtained by mapping local fluctuations in area fraction and density of pinned particles onto the bulk diagram of Fig. \ref{figEntropyMap}, while the latter is immediately accessible from the particle coordinates. Most importantly, $\tilde{s}$ is completely blind to the presence of pinned particles, while $s^{\rm loc}$ explicitly depends on it.}

\section{Equilibrium behaviour}
\label{secEquilibrium}
\subsection{Phase diagram}

\begin{figure}[t]
	\includegraphics{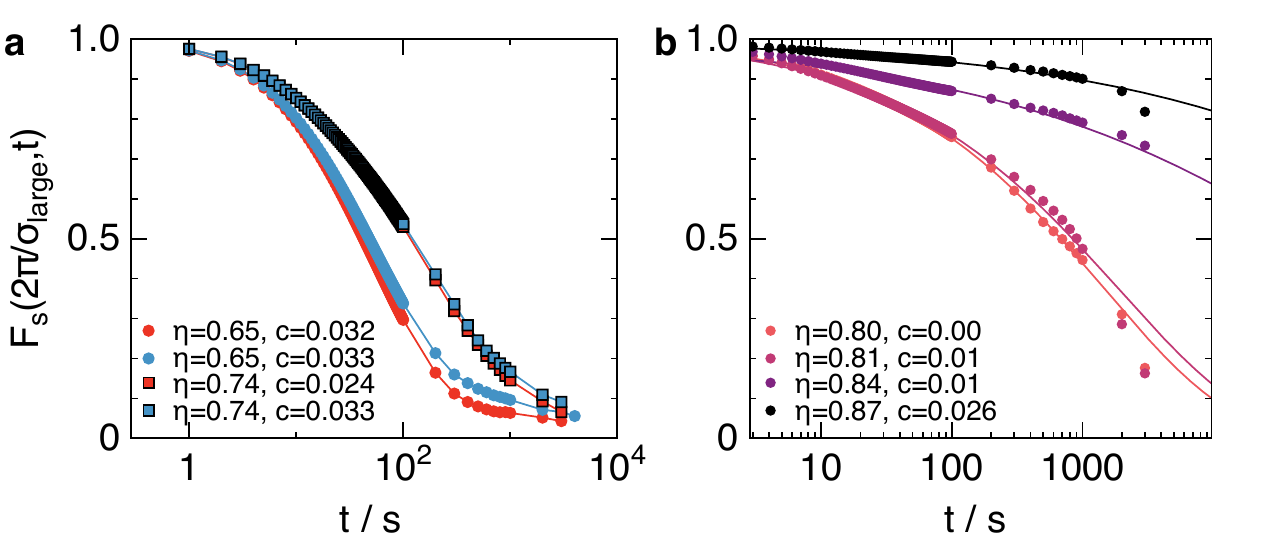}
	\caption{\ian{Self part of the Intermediate Scattering Function (ISF) calculated from experiments. (a) For area fractions below $\eta\approx0.80$ we are able to observe full relaxation and differences in lower (red) and higher (blue) pinning concentrations are measurable. (b) For area fractions $\eta\geq0.80$ the accessible time-window is insufficient to observe full relaxation and the system is a glass. Lines are stretched exponential fits from which relaxation times are \ft{estimated}. }}
	\label{figISFs}
\end{figure}

The (metastable) equilibrium phase behaviour of the liquid is inferred from numerical simulations of $N=6000$ particles, \ian{with pinned concentrations in the range $c\in[0,0.16]$,} \ft{in 5 independent runs}. We compute the structural relaxation time $\tau_{\alpha}$ from the self part of the intermediate scattering function \ft{$F_s (|\vec{q}|,t)$} defined as
\ft{\begin{equation}
	F_s(q,t)|=\langle e^{ i\vec{q}\cdot(\vec{r}(t)-\vec{r}(t_0))}\rangle_{t_0}
\end{equation}
where we take $q=|\vec{q}|=2\pi/\sigma_{\rm large}$ and the average is performed on equilibrium (or stationary, in the case of the experiments) trajectories. In Fig.~\ref{figISFs} we show the correlation functions} \ian{obtained from}  \ft{stationary experiments} \ian{at a range of area fractions and pin concentrations. For area fractions $\eta \lesssim 0.8$, full relaxation is observed within the experimental time window for all pinning concentrations studied [Fig.~\ref{figISFs}(a)]. When the area fraction exceeds $\eta = 0.8$, however, full relaxation does not occur on the experimental timescale [Fig.~\ref{figISFs} (b)], the sample is not equilibrated and must be considered a glass, similarly to previous work \cite{illing2017mermin,gokhale2014, nagamanasa2015, pastore2015connecting}. } 

\begin{figure}
	\centering
	\includegraphics[width=\textwidth]{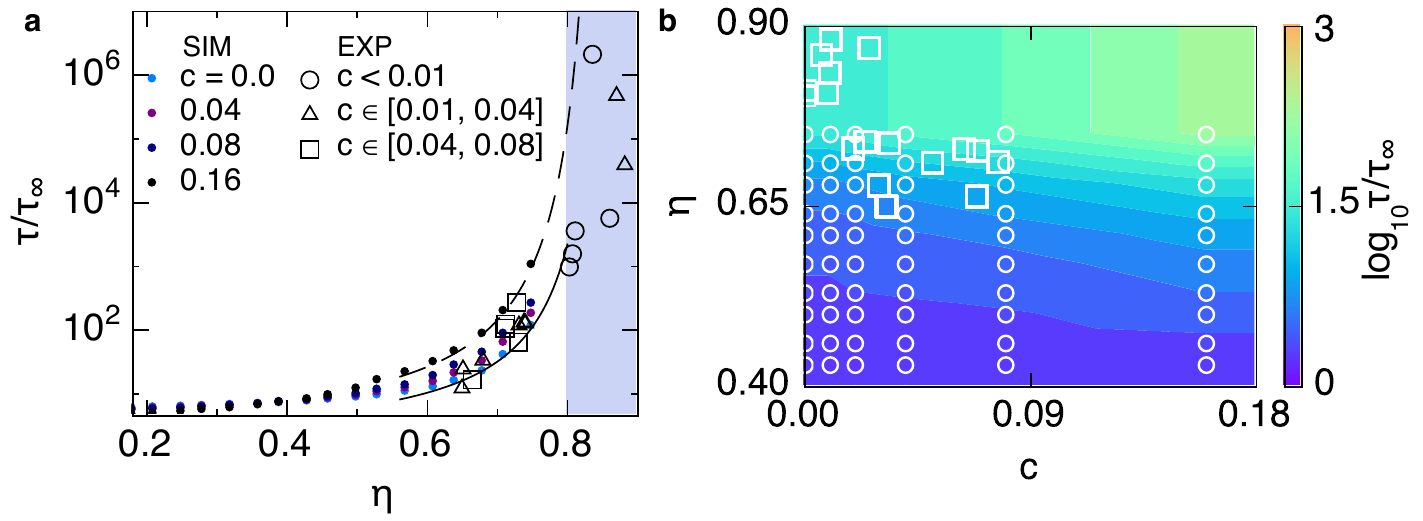}
	\caption{ (a) \ft{Relaxation times measured in simulation (coloured dots) and experiment (empty symbols) as a function of the total area fraction $\eta$. The shaded area indicates the region where full relaxation cannot be observed in the available observation time, and where the reported relaxation times are estimated from ageing samples. The continuous and dashed lines are $\log \tau/\tau_\infty = D/(\eta-\eta_0)$ fits to the $c=0$ and $c=0.16$ simulation datasets. (b) Dynamical phase diagram at high area fractions: the experiments (squares) span a wide range of area fractions $\eta$ and pinning concentration $c$. Numerical simulations (circles) provide equilibrium measurements of the relaxation times. We fill the represented phase diagram through numerical interpolation, and a logarithmic colormap for $\log_{10} (\tau_\alpha/\tau_0)$. }}
	\label{figangell}
\end{figure}

\ian{The relaxation time, $\tau_{\alpha}$, is extracted by fitting the intermediate scattering function decay with $F_s(q,t)= a \exp (-(t/\tau_{\alpha})^b) $, the Kohlrausch-Williams-Watts law with $0\leq a\leq 1$ and $0< b\leq 1$. Figure \ref{figangell}(a) shows $\tau_{\alpha}$ scaled by the low density limit $\tau_0=\lim_{\eta\rightarrow 0} \tau_\alpha$, calculated from both experiments and simulations as a function of area fraction. For $\eta < 0.8$ we find good agreement between experiment and simulation. Relaxation time is found to be more strongly dependent on $\eta$ than pinning concentration, with small changes in area fraction within the supercooled regime having a much larger effect on $\tau_{\alpha}$ than small changes in $c$. Simulations (small solid points) do reveal a upwards shift, representing slower dynamics, as $c$ is increased from zero. However, due to uncertainties in $\eta$ and $c$ and the difficulty in obtaining good experimental statistics, this relationship is not so clearly evident in the experimental data. For $\eta > 0.8$ (shaded region) we have no simulations, only glassy experimental data, showing large fluctuations in measured $\tau_\alpha$, characteristic of a kinetically trapped, out-of-equilibrium sample.} 



\ian{The relationship between area fraction, pin concentration and structural relaxation time is illustrated} in Fig. \ref{figangell}(b) \ian{where} we \ian{interpolate a colormap representing} the logarithm of the relaxation times \ft{obtained from simulations. Open white circles represent the locations of the simulations from which the interpolated map is constructed, while open white squares refer to the experimental conditions. This map reiterates the data shown in Figs.~\ref{figangell}(a).} As expected, \ian{an} increase \ian{in $c$} corresponds to a progressive slowing down of the dynamics for a given area fraction. \ian{This is most} pronounced for \ian{low} area fractions between $0.5<\eta<0.65$ \ian{corresponding to very fluid samples. From simulations we learn that for, \emph{e.g.}, $\eta \approx 0.68$, one needs to immobilise $\sim 15\%$ of all particles to observe an order of magnitude decrease in $\tau_{\alpha}$. The maximum quasi-static pinning concentration obtained in experiments in this $\eta$ regime is $c \sim 0.08$, for which the simulations show we expect only a small slowing of dynamics. Thus, it is perhaps not surprising that it is more challenging to resolve pin-induced slowing down in our experiments.} High area fractions \ian{are not sampled} in equilibrium, so we do not exceed $\eta=0.78$ in equilibrium \ian{simulations}. \ft{As mentioned above, experiments are performed at higher area fractions, but do not fully relax within the observation time} \ian{(Fig.~\ref{figISFs})}.

\begin{figure}
	\centering
	\includegraphics[scale=1]{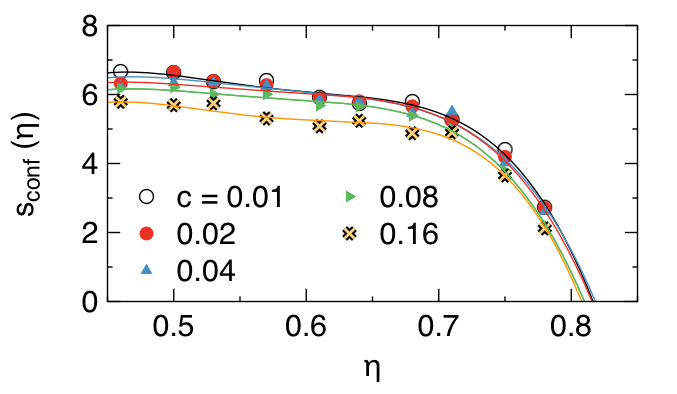}
	\caption{Configurational entropy as a function of area fraction, $\eta$, for different pinning concentrations, $c$, as obtained from the method \ft{described in the Appendix for equilibrium molecular dynamics simulations}. Lines are polynomial fits to guide the eye. Throghout, the Boltzmann constant is set to $k_B=1$.}
	\label{figConfEnt}
\end{figure}

\begin{figure}
	\centering
	\includegraphics{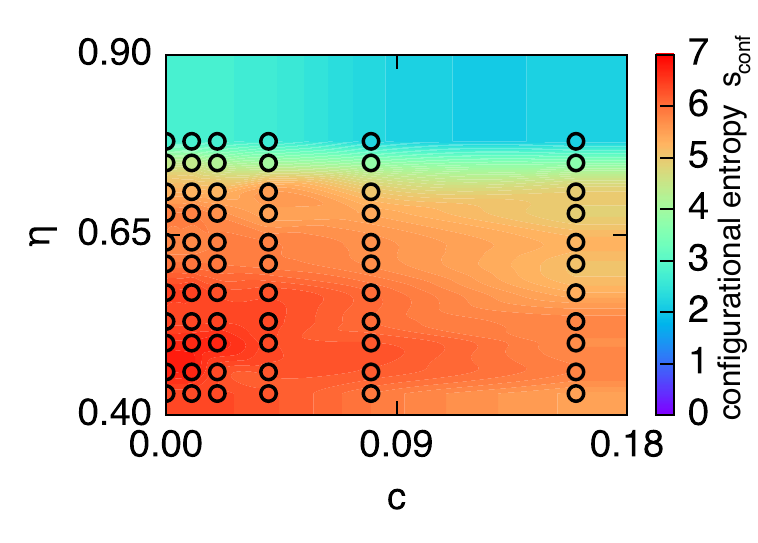}
	\caption{Color plot of the configurational entropy $\sconf(\eta,c)$ \ft{as obtained from numerical simulations}. Circles indicate the points evaluated in simulation; the level curves are obtained by quadratic interpolation and show that pinning has \ian{only a} moderate effect on the configurational entropy of the system even at relatively high concentrations of pins.}
	\label{figEntropyMap}
\end{figure}

From the equilibrium simulations we also obtain the configurational entropy \ft{following the thermodynamic integration method detailed in the Appendix} \ian{as shown in} Fig.~\ref{figConfEnt}. \ian{This method} requires \ian{calculating the vibrational entropy through} the application of additional harmonic potentials \ian{which} is unfeasible in experiments under the current protocol, and \ian{so} we limit ourselves to \ian{employing these} numerical results to guide our analysis \ian{and interpretation of the experiments.} \ian{As with relaxation time,} \ft{the numerical results display} a transition to lower entropy \ft{that is smooth and continuous} as the concentration of pins increases. \ian{However,} as in the case of the relaxation times, small variations in the area fraction typically produce larger effects than comparable variations in the pinning concentration. \ian{This is particularly evident in the color map shown in Fig.~\ref{figEntropyMap}.}

We notice that the progressive arrest of the system due to pinning corresponds to both an increase of the structural relaxation time and the reduction of entropy for the bulk system. While the effect of pinning on the structural relaxation time (and hence the mobility) is obvious, \ian{its} effect on configurational entropy is non-trivial. A conventional picture relates the pinning process to a progressive simplification of the configurational space, with entire families of configurations \ian{becoming} excluded from the possible ensemble of configurations that the system can assume. In this sense, as pinning progresses, the number of alternative configurations decreases and so does the entropy. However, it is not clear whether such a process is accompanied by structural \ian{changes} at the local level. In the following we show that, within the limits of our approach, no significant restructuring occurs when pinning progresses.

\subsection{\ft{Local structure and mobility}}
\ian{We characterise} the structural features of the system \ian{using} local area fraction, \ian{which is} easy to measure \ian{in} both simulations and experiments, and the \ft{local entropy $s^{\rm loc}$, Eq. \ref{eq:sloc}, which is read from the interpolated phase diagram in Fig. \ref{figEntropyMap} for any value of the local pinning fraction $c$ and the local area fraction.}

\begin{figure}[t]
	\centering
	\includegraphics[width=\textwidth]{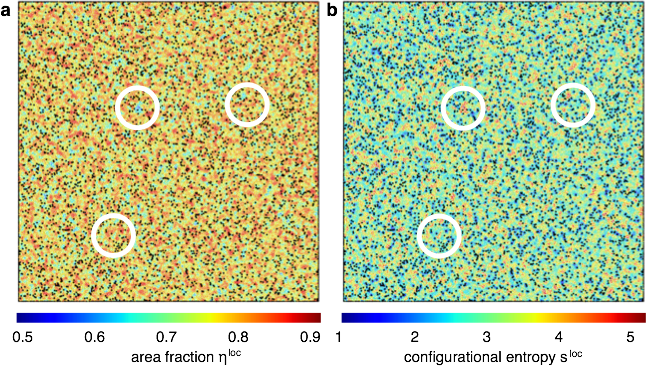}
	\caption{Maps of local area fraction (left) and configurational entropy (right) for a numerical simulation of total area fraction $\eta=0.74$ at pinning $c=0.08$. Black dots indicate the positions of pinned particles. White circles \ian{highlight} regions demonstrating \ian{similar} spatial patterns \ian{in both} representations.}
	\label{figVoroMaps}
\end{figure}

Figure~\ref{figVoroMaps} shows colormaps \ian{representing local} area fraction, $\etl$, and entropy, $s^{\rm loc}$, \ian{in} a simulation of $60000$ particles  at $\eta=0.74$ with an intermediate pinning concentration $c=0.08$. Spatial correlations between the two quantities are immediately evident and we highlight a few \ian{highly similar regions}. It appears, therefore, that area fraction fluctuations essentially \ian{predict} the shape of low and high entropy regions, regardless of the local fluctuations of the pinning concentration. \ft{This was anticipated from the bulk phase diagram of Fig.~\ref{figEntropyMap} which showed that variations in the area fraction affect the entropy more strongly than variations in the pinning fraction.}

\begin{figure}
	\centering
	\includegraphics{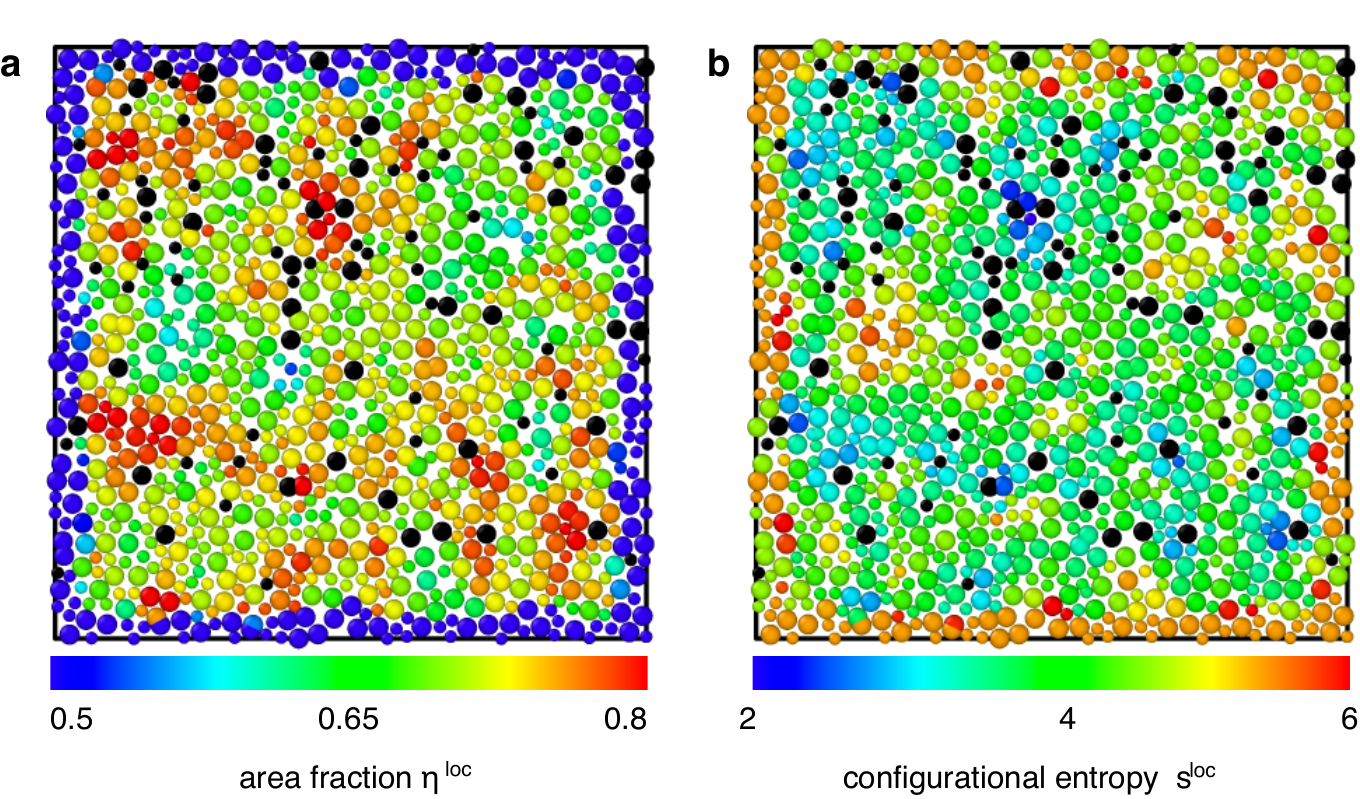}
	\caption{\ft{Experimental configuration at $\eta=0.715$ with pinning concentration $c=0.08$. As in Fig.~\ref{figVoroMaps}, we plot the local area fraction from the Voronoi tessellation (a) and the interpolated local entropy $s^{\rm loc}$ (b) as mapped through the diagram in Fig.~\ref{figEntropyMap}. High area fraction regions are typically mapped into low entropy regions. Particles at the borders of the field of view are not physically relevant, as their Voronoi surfaces cannot be estimated and hence their area fractions are very low.}} 
	\label{figExpExample}
\end{figure}

\ft{Such color coding of the particle coordinates can also be applied to the experimental datasets, assuming that the phase diagram in Fig.~\ref{figEntropyMap} is valid for the experimental conditions. In Fig.~\ref{figExpExample}  we consider an experimental sample at $\eta=0.715$ with an intermediate value of pinning fraction, c=0.08. In Fig.~\ref{figExpExample} (a) and (b) we color-code the tracked particle coordinates according to $\etl$ and $s^{\rm loc}$ as read from Fig.~\ref{figEntropyMap}. We \ian{again highlight} that low entropy regions correspond to high area fraction regions, with the \textit{caveat} that the Voronoi cells (and the derived local quantities) of  particles at the edges of the imaged region cannot be reliably computed.} 

\ft{In order to perform a quantitative analysis of  spatial correlations of the local area fraction, entropy and mobility fields, we restrict ourselves to inner regions, far from the edges. We define the mobility as the distance travelled by a particle in a time interval $\Delta t\sim \tau_\alpha$}

\begin{equation}
m^i (t_x)=|\vec{r}_i (t_x+\Delta t)-\vec{r}_i (t_x)|.
\label{eq:mobility}
\end{equation}

\begin{figure}
	\centering
	\includegraphics{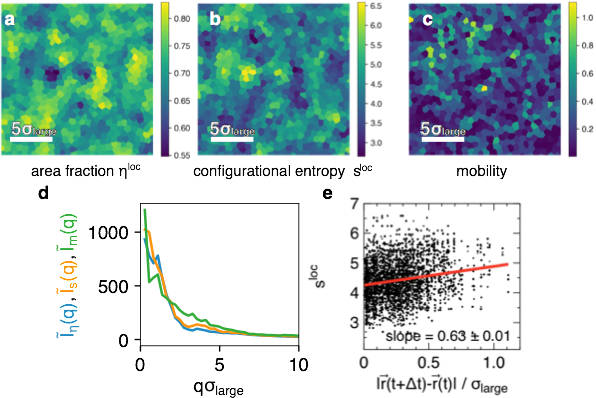}
	\caption{\ft{Spatial analysis of an experimental sample at $\eta=0.715$ with a pinning concentration of $c=0.08$.  In (a) and (b) and (c) the local area fraction, the interpolated local entropy and the local mobility in the interior regions of the sample (\textit{i.e.} excluding the borders) are mapped onto fine two-dimensional grids of $256\times 256$ cells. In (d) we plot the radially averaged power spectra $\tilde{I}_\eta(q)$ (blue), $\tilde{I}_{s}(q)$ (orange), $\tilde{I}_{m}(q)$ (green) of the scaled Fourier transforms of (a), (b) and (c).}} 
	\label{figExpExample3}
\end{figure}

\ft{Through mapping the local area fraction, local entropy and mobility onto a fine two-dimensional grid, as in Fig.~\ref{figExpExample3} (a), (b) and (c), we obtain images $I_{\eta}$, $I_s$, and $I_m$ of local area fraction, entropy and mobility respectively. We compute the rescaled two-dimensional Fourier transforms as $\tilde{I}_x= \mathcal{F}[(I_x-\min I_x)/(\max I_x-\ \min I_x)]$ for $x:\eta,s, m$  where the $\mathcal{F}$ operator represents the discrete Fourier transform. In Fig.~\ref{figExpExample3} (d) we plot the radially averaged absolute value (or power spectrum) of $\tilde{I}_x (q)$ and find that the power spectra of local area fraction and entropy are overlapped in a wide range of the wave-vector $q$, displaying similar features at particular modes, such as a peak at $q\sigma_{\rm large}\approx 3.6$, while the mobility spectrum is globally blind to such features. In Fig.~\ref{figExpExample3} (e) we directly plot the inferred local entropy against the measured mobility, revealing a weak positive correlation between the two --- larger displacements tend to be observed in high entropy regions. This analysis suggests (1) that features present in the spatial fluctuations of the local area fraction field are reproduced in the local entropy field and (2) that weak positive correlations exist between the local entropy and the measured particle mobility. The weak correlation between a local static measure such as the area fraction (or the interpolated local entropy) and the mobility is compatible with the finding that the correlation between dynamical heterogeneities and local structure is highly system-dependent \cite{starr2002,hocky2014,pastore2017}.}  

\begin{figure}[t]
	\centering
	\includegraphics{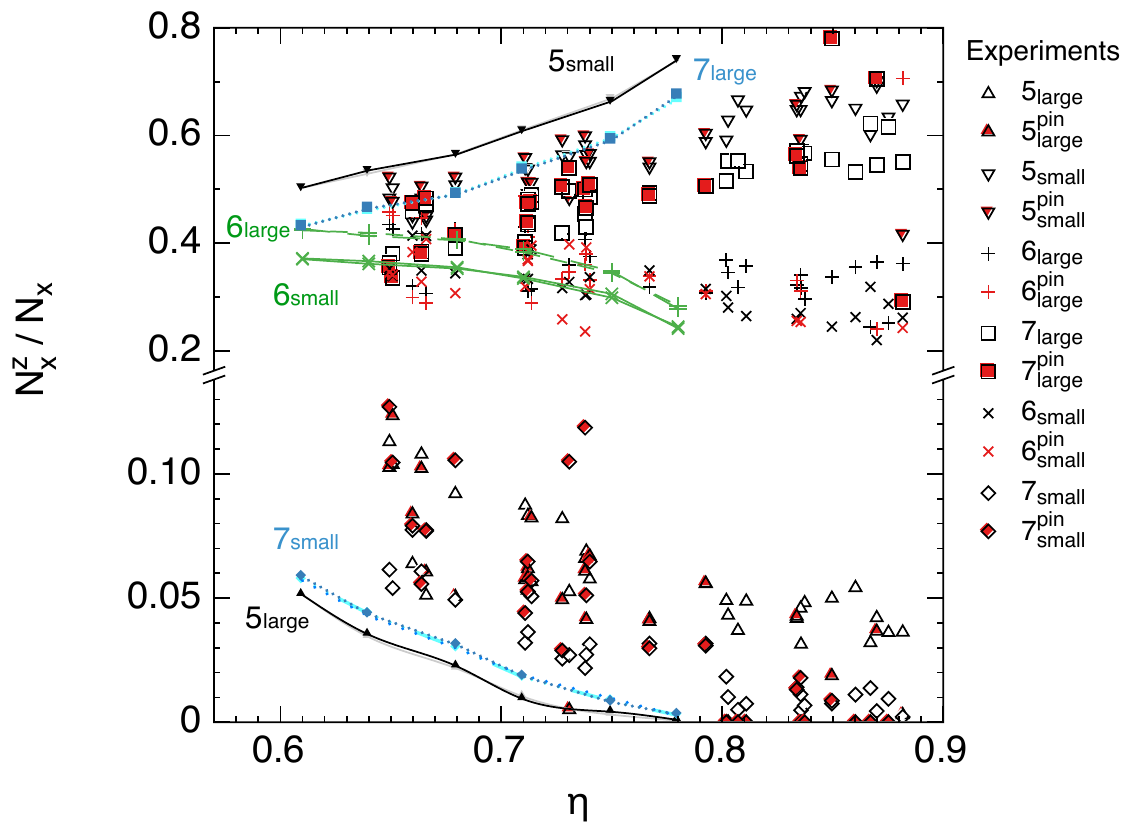}
	\caption{\ft{Comparison between the particle coordination in experiments and simulations. We plot the fraction of particles of type $x$ that have coordination $z$. Empty symbols represent experimental data for all large/small particles. Filled red symbols correspond to pinned particles. Connected symbols are results from the simulations: coordination $z=5$ particles are in black ($\blacktriangle$: large, $\blacktriangledown$: small), $z=6$ in green ($+$: large, $\times$: small) and $z=7$ ($\blacksquare$: large, $\blacklozenge$: small) are in light blue.}}
	\label{figCoordination}
\end{figure}

\ft{To complete the picture of static correlations we compare the coordination number  around small and large particles in experiments and simulations with stationary pinning, Fig.~\ref{figCoordination}. We compile all experimental data regardless of the pinning concentration and plot the fraction of particles of type $x$ that have coordination $z$, $N_x^z/N_x$, with particular focus on $z=5,6,7$ (pentagonal, hexagonal and heptagonal local order respectively).} \ft{In numerical simulations we pin particles from well equilibrated configurations for every total area fraction. Therefore, we expect the local order to be solely determined by the total area fraction. We indeed observe that the fraction of particles with a given coordination does not vary between $c=0$ and $c=0.16$. Hence, we plot $N_x^z/N_x$ in Fig.~\ref{figCoordination} from the simulations only as a function of $\eta$ in order to guide the interpretation of the experimental measurements.}

\ft{Both simulations and experiments show that the fraction of small particles with coordination $z=5$ and the fraction of large particles with $z=7$ increase with increasing total area fraction, while the reverse occurs for small particles with $z=7$ and large particles with $z=5$. The fraction of particles with coordination $z=6$ decreases as the area fraction increases, for both large and small particles, both in simulations and in experiments. For $\eta\geq 0.8$ the experiments are not equilibrated and we indeed observe (within the scatter of the data) that the fraction of $z$-coordinated particles reaches a plateau. In previous analysis of a soft-core two-dimensional glass former \cite{eckmann2008}, the large/small particles in pentagonal/heptagonal Voronoi cells were interpreted as ``liquid-quasispecies'' while large/small particles in heptagonal/pentagonal cells were seen as ``glasslike quasispecies''. The evolution of the populations of the $z$-coordinated particles is in agreement with these trends and the plateau in the experiments for $\eta>0.80$ is another signature of the nonequilibrium nature of the high area fraction samples.} \ft{In the case of the experiment, we also analyse the subpopulation of pinned particles in order to identify eventual systematic differences between them and the total population. We do not find any significant discrepancies between the two, supporting the idea that adsorption and hence pinning in experiment are not determined by the nature of the local order.}

\ft{In conclusion, we have shown, from numerical simulations, that the configurational entropy of the liquid is reduced as the area fraction is increased and the relaxation times increase. At the local level, we are able to map local area fraction and local pinning concentration to entropy fluctuations. This mapping is such that the spatial extent of entropy fluctuations is strongly correlated with area fraction fluctuations but only weakly correlated to mobility fluctuations. As the bulk configurational entropy decreases with increasing area fraction, we have confirmation from both numerical simulations and the experiments that local hexagonal order is suppressed while large particles become mostly $7$-coordinated and small particles $5$-coordinated, regardless of whether or not a particle is pinned.}

\section{Non-equilibrium growth of the pinning concentration}
\label{secNoneq}
\ian{In an experimental sample, particles become irreversibly pinned as time proceeds, so that, with suitable preparation and observation time, it is possible to measure a global increase in the pinning concentration. Here we report particular experimental instances in which significant increase in $c$ is observed. The system here is out of equilibrium and therefore represents a fundamentally different pinning protocol to that discussed in the previous section.} The system has \ian{insufficient} time to \ian{relax and} de-correlate at \ian{each} successive \ian{step in} pinning concentration. The correlation time increases \ian{due to} the progressive increase in $c$ and the system \ian{traverses} the \ft{unpinned liquid/pinned} glass crossover.

\begin{figure}[t]
	\centering
	\includegraphics[scale=0.6]{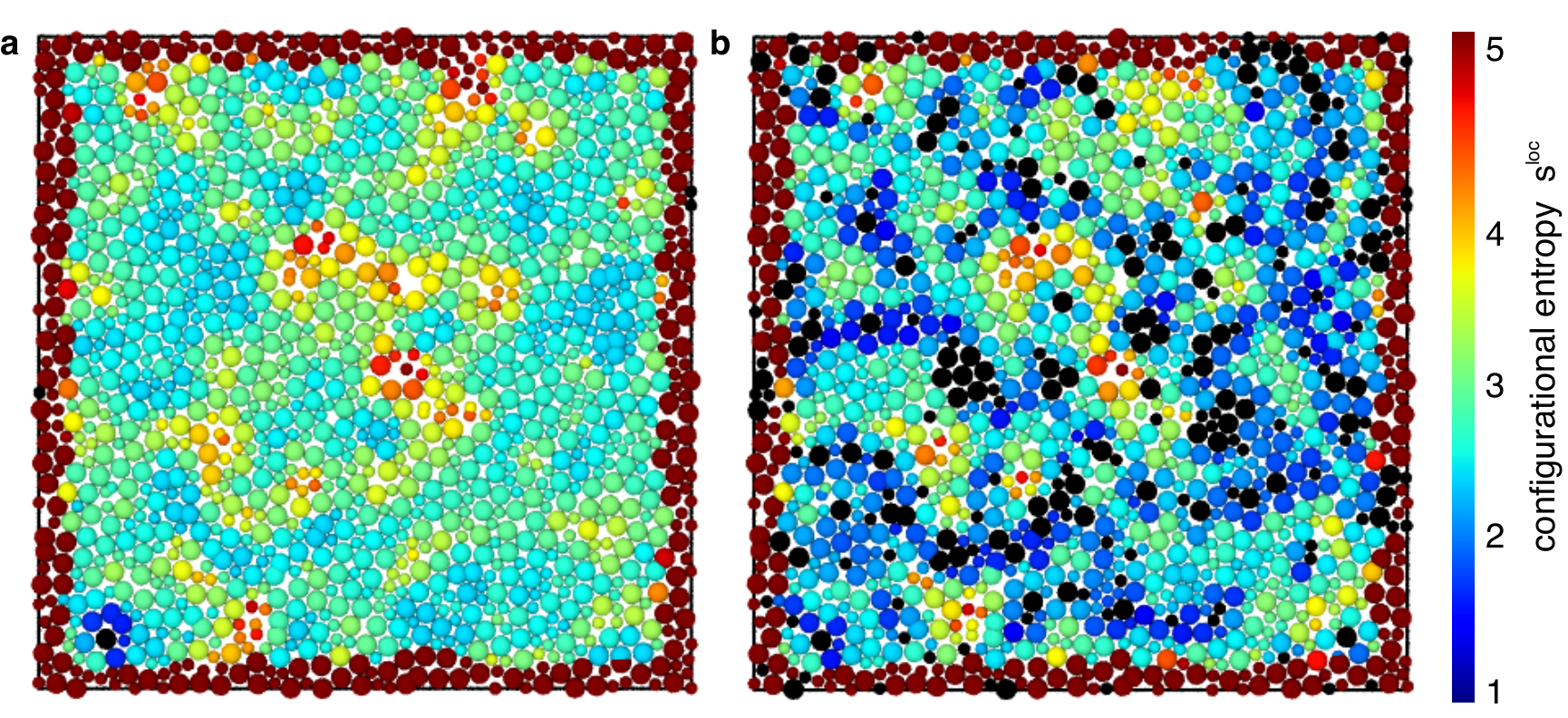}
	\caption{The increase in \ian{pinning concentration} is mapped \ian{onto} changes \ian{in} the local configurational entropy $s^{\rm loc}$ in an experiment with area fraction $\eta=0.85$. \ian{Particle colour represents local configurational entropy as indicated in the colour bar. Black particles are identified as immobile.} (a) At time $t_1 = 1000 \; \mathrm{s}$, few particles are pinned and the areas of low configurational entropy are limited (lower left corner). (b) \ian{By} a later time $t_2=2000 \; \mathrm{s}$ the pinning concentration \ian{has risen} to $c=0.15$ and more extended regions of low entropy appear as a direct consequence.}
	\label{figExpEntropy}
\end{figure}

Using the interpolated configurational entropy $s^{\rm loc}$, Fig.~\ref{figExpEntropy} \ian{shows} that local entropy \ft{appears} strongly correlated to the increase \ian{in} pinning concentration in experiments. \ft{However, this is a consequence of the} interpolation procedure, which takes into account local area fraction and pinning concentration, shifting the local entropy to lower values when \ian{sufficiently} large pinning concentrations are measured. \ian{Furthermore,} pinned particles freeze-in local area fraction fluctuations and these are reflected in the local \ian{entropy} fluctuations.

\ian{Are changes in local structure observed as $c$ increases? To address this question we \ft{make use of} the} definition of two-body excess entropy. We analyse the progressive immobilisation of the system with two parameters: the local two-body entropy, Eq.~\ref{eq:stwobody}, and the root square displacement of the particles from reference configurations. We measure the local two body entropy, $\tilde{s}$, at time $t_x=t_0,\dots t_n$ and compare the \ian{resulting two-body excess entropy} colormaps with \ian{those} obtained from the \ian{single particle} mobility defined in Eq.~\ref{eq:mobility}, where $\Delta t\approx 200 s$.

\begin{figure}[t]
	\centering
	\includegraphics[width=\textwidth]{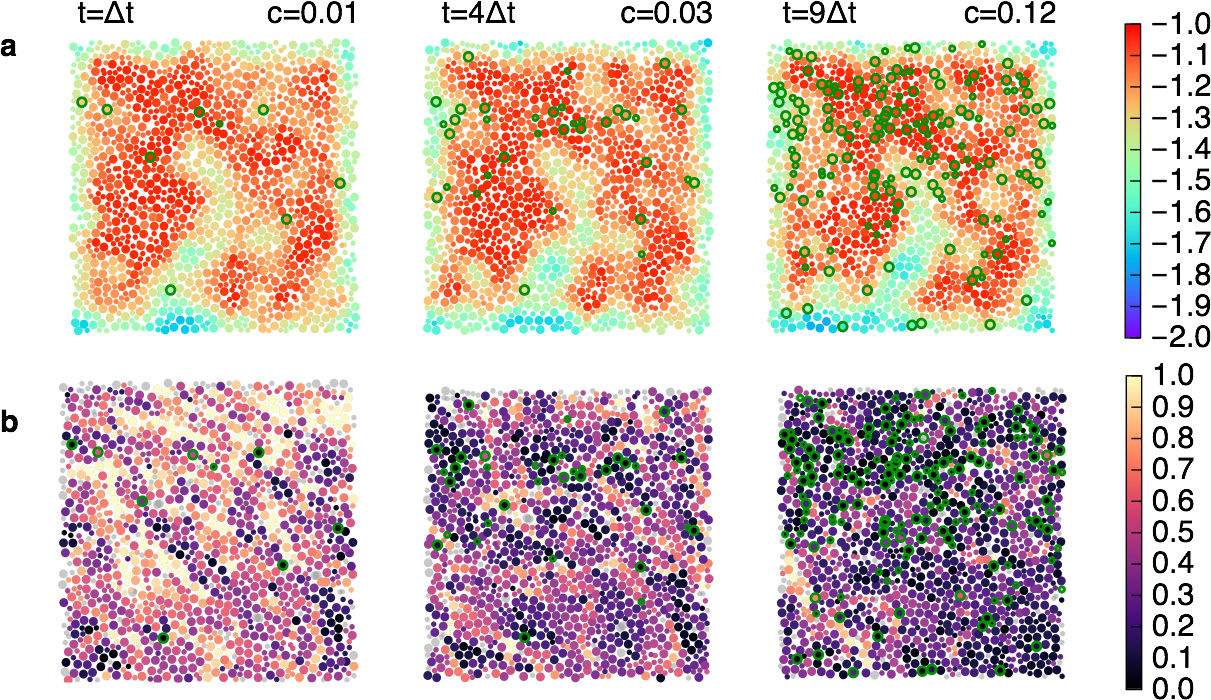}
	\caption{Time evolution of \ian{experimentally measured} (a) local two-body excess entropy and (b) mobility $m(t)$ in units of $\sigmal$ for $t=1\Delta t, 4\Delta t, 9\Delta t$ at $\eta=0.83$. Green circles indicate pinned particles at concentration $c(t)$. In grey we plot particles \ft{for which mobility cannot be computed as they are not identified both in the initial and final frame}.}
	\label{figTimeEv}
\end{figure}

In Fig.\ref{figTimeEv} we show \ian{two-body excess entropy and mobility} maps for a system with area fraction of $\eta=0.83$. A spatial correlation between \ian{these quantities is not immediately evident, except for} some features at early times (\emph{e.g.} the low entropy region in middle of the sample corresponds to a compact region of low mobility). As \ian{more particles become pinned}, the mobility field \ian{appears to} gradually \ian{flatten}, \ian{suggesting that dynamical heterogeneity is suppressed}, while \ian{little change is observed in the entropy field.}

\begin{figure}[hbt]
	\centering
	\includegraphics[width=\textwidth]{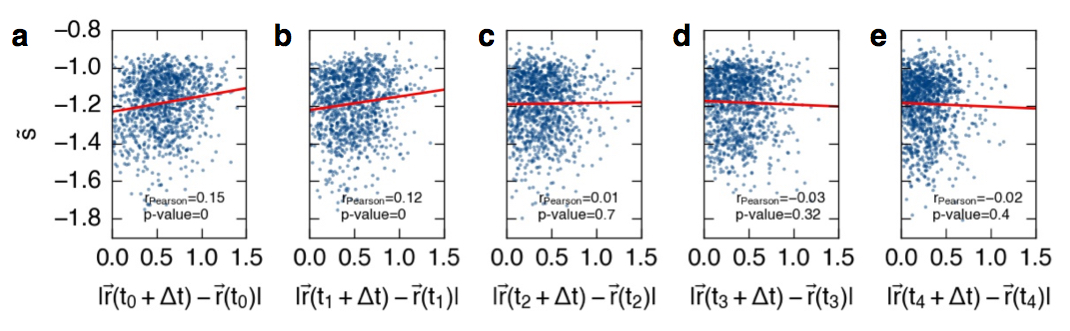}
	\caption{Correlation (or lack \ian{thereof}) between the local entropy $\tilde{s}$ and the mobility as time evolves and the concentration of pinned particles increases: (a) $c=0.01$, (b) $c=0.018$,(c)$c=0.026$,(d) $c=0.037$, (e) $c=0.075$. }
	\label{figMobScorrelation}
\end{figure}

In order to substantiate \ian{the above} statements more quantitatively, we plot the correlations between entropy and mobility \ian{as pinning fraction increases,} Fig.\ref{figMobScorrelation}. Here we see that if a weak correlation is present at early times, this is not the case at later times and higher pinning concentrations. In particular, the very large increase of the p-value indicates that \ft{the null hypothesis that the correlation coefficient is zero is increasingly plausible as time evolves, and that} any eventual correlation at high pinning concentration is spurious. \ft{In these conditions,} the local entropy (\ian{and} hence \ian{any change in structure}) is blind to the dramatic decrease of mobility induced by pinning. In this sense, the changes in the pinning field do not correspond to structural changes significant enough to be identified by the local two-body excess entropy.

\section{Conclusions}
In this work we present a novel \ian{method} to study dense two-dimensional suspensions of hard colloids under the effect of a field of immobilised particles. We have shown that it is possible to realise systems that are at steady state for several hours, allowing the measurement of local fluctuations \ian{in} area fraction. Using computer simulations, we have interpreted our data showing that there is a strong anti-correlation between the spatial fluctuations in area fraction and the reduction of local entropy, according to two alternative measures of local entropy. The presence of pinned particles does not appear to change the structure of the liquid but, as expected, it reduces the bulk entropy \ian{by restricting the} accessible configurational space. \ft{Structural changes in the particle coordination are solely determined by the increase in area fraction}. \ft{With the concentration of pinned particles kept constant, local entropy fluctuations are weakly positively correlated with the reduced mobility of the particles.}

On longer time-scales, our \ian{protocol} allows us to \ian{observe} the response of the system to a monotonic increase of the concentration of pinned particles. We attempted to identify any predictive power in the local fluctuations of the two-body excess entropy, in particular with respect to the mobility of the system. The available data \ian{suggests} that local entropy fluctuations at time $t_0$ are weakly correlated with the local mobility at immediately subsequent times \ian{but are} uncorrelated as time \ian{progresses} and pinning concentration \ian{increases}. While the progressive \ian{increase in} pinning has deep consequences on the global scale, \ian{its local consequences are} trivial. In particular it does not appear to affect the structure, as expected from theory \cite{cammarota2011}. 

Our experimental method is competitive with respect to alternative pinning techniques such as optical tweezers as it allows us to perform large scale pinning at a very limited cost, \ian{providing} detailed \ian{particle-level structural and dynamic} information. However, \ian{the} drawback of \ian{approaching this problem} in two-dimensions is that no sharp transition \ian{occurs} and only \ian{a weak} crossover between an unpinned and a pinned regime can be studied. Fluctuations \ian{in} order parameters such as the patch-entropy of Sausset and Levine \cite{sausset2011} or a measure of the degree of local crystalline order \cite{russo2015} could help to characterise the continuous emergence of the arrested phase. Considering local structural signatures based on higher order expansions of the excess entropy accounting for multiparticle correlations \cite{banerjee2016effect} could \ian{potentially reveal} more subtle structural changes triggered by pinning, relating higher order static correlations to the dynamics \cite{royall2015physrep}.

\ian{Overall, our extensive experimental and simulation study shows that the effects of random pinning in two dimensional glass formers is very subtle. Thoughtful and careful analysis is required to disentangle the effects of system density and pinning concentration on the observed structure and dynamics. We have, however, revealed various correlative relationships which may serve as a roadmap for more targeted future studies in two dimensional systems under random pinning.}

\ack{IW, FT, JEH, PC and CPR acknowledge the support of the European Research Council (ERC Consolidator Grant NANOPRS, Project No. 617266). CPR also acknowledges the support of the Royal Society and Kyoto University SPIRITS.}

\appendix
\section{Configurational entropy}

We consider the system as a binary mixture of ideal hard spheres, for which we know the equation of state $P(\eta,c)$ from the Molecular Dynamics calculations. We compute the configurational entropy from the difference between the total entropy and the vibrational part. The total entropy is computed with reference to an ideal gas filling a box with a prescribed concentration of pins $c$ that represent inaccessible regions. \ft{Throughout, we very closely follow the approach of Angelani and Foffi \cite{Angelani2005,angelani2007}, who originally illustrated the method for a binary mixture of hard spheres. We refer to their work for a more detailed discussion of the validity and the limitations of such numerical technique. Here, we underline that this route constitutes one of the possible methods to compute the configurational entropy of supercooled liquids. In fact, that recent calculations \cite{berthier2017} show that this method is in qualitative agreement with alternative methods, despite a systematic overestimation of the number of distinct metabasins. }
\newline 

We start writing the total entropy as the sum of an ideal and excess contribution
\begin{equation}
	\Stot (\rho,c)=\Sideal(\rho)+\Sexc(\rho,c)
\end{equation}

We need to take care of the distinction between pinned and unpinned particles. The number of mobile (unpinned) particles is $M=N(1-c)$ where $N$ is the total number of particles and $c$ the pinning concentration. The density of non-pinned particles $\rho_M$ is defined as number of mobile particles $M$ divided by the accessible area $A_a=A-N ca$, where $A$ is the total area of the box and for our equimolar binary mixture $a=\pi(\sigmal^2+\sigmas^2)/8=\eta/\rho$, that is 

\begin{equation}
\rho_M=M/A_a=\rho (1-c)/(1-c\eta).
\end{equation}

The configurational entropy for $M$ ideal gas particles at density $\rho_M$ is given by 
\begin{eqnarray}
	\Sideal (\rho_M,c)& =M(2-\ln \rho_M +\ln 2 -2 \ln \lambda)\\
	& = (1-c) N \left(2-\ln\rho \dfrac{(1-c)}{1-c\eta}+\ln 2\right),
\end{eqnarray}
where the $\log 2$ term accounts for the entropy of mixing and $\lambda$ is the thermal de Broglie wavelength $\lambda = \sqrt{2\pi \beta\hbar^2/m}$ that, for simplicity, we set to $\lambda=1$ .
The excess part is computed from the equation of state 
\begin{equation}
	\Sexc (\rho_M,c) =-\beta M \int_0^{\rho_M}d\rho_M'\frac{P_{\rm exc}(\rho_M, c)}{\rho_M'^2},
	\end{equation}
which, in terms of the total density, reads
\begin{equation}
	\Sexc(\rho,c)=-\beta M \int_0^{\rho} \frac{P_{\rm exc}(\rho,c)}{\rho'^2} \frac{1-c}{(1-\rho' ca)^2}d\rho'.
\end{equation}

From the sum of the ideal and excess contributions we subtract the vibrational part. \ft{Following closely the approach detailed in \cite{Angelani2005}, the vibrational contribution is obtained through a thermodynamic integration approach. The original Hamiltonian is modified so that we obtain an augmented Hamiltonian}
\begin{figure}[t]
	\centering
	\includegraphics{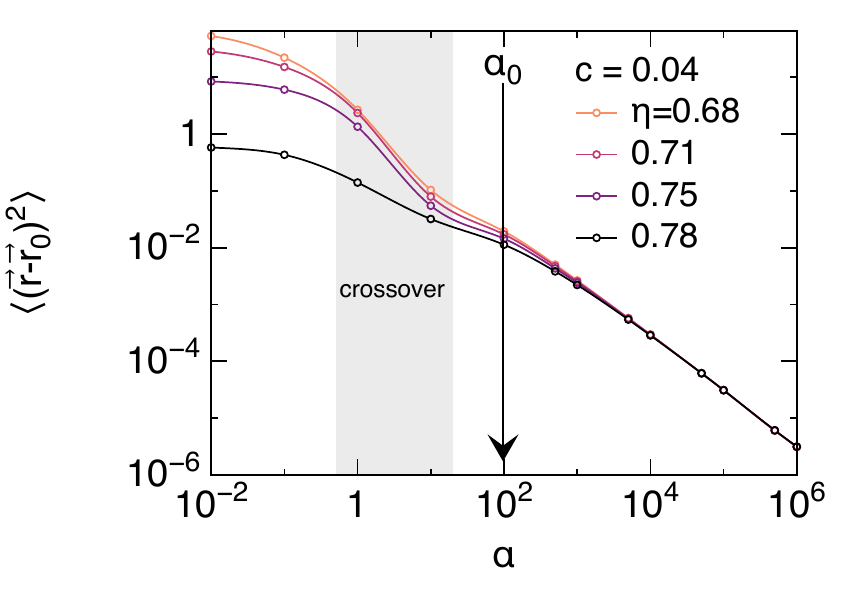}
	\caption{Mean squared displacement from a reference configuration $\vec{r}_0$ for a system with pinning concentration $c=0.04$ and several values of the area fraction $\eta$. We choose $\alpha_0 =100$  indicated by an arrow), just above the crossover region where the particles start to move far from their original positions. }
	\label{figexampleMSDvsAlpha}
\end{figure}
\begin{equation}
	\beta H'=\beta H+\frac{1}{2}\alpha M(\vec{r}-\vec{r}_0)^2,
\end{equation}
\ft{where the second term represents springs of strengths $\alpha$ coupled to a reference configuration $\vec{r}_0$, chosen among equilibrium configurations. The exact same argument as in \cite{Angelani2005} can be followed, the main difference being the dimensionality of our system. Hence, the free energies $F(\alpha_0), F(\alpha_{\infty})$ of two systems with different spring constants $\alpha_0,\alpha_\infty$ are related by}
\begin{equation}
\beta F(\alpha_\infty)=\beta F(\alpha_0)+\int_{\alpha_0}^{\alpha_\infty}d\alpha'/2 \left\langle \sum_{i=1}^{M}(\vec{r}-\vec{r}_0)^2\right\rangle_{\alpha'}, 
\end{equation}
\ft{where $\langle\dots\rangle_{\alpha'}$ is the canonical average for a given $\alpha'$. For $\alpha_{\infty}\rightarrow\infty$ (very hard confining potential) the particles cannot escape and the free energy is $\beta F(\alpha_\infty)=2 M \ln \lambda+\beta E_0+M\ln(\alpha_\infty/2\pi)$, where $E_0$ is the energy of the reference configuration $\vec{r}_0$. Writing the free energy as a sum of a potential and an entropic term $\beta F(\alpha_{\infty})+ M -S_{\rm vib}$ it is possible to write down the vibrational entropy as:}
\ft{
\begin{equation}
	\Svib=M\int_{\alpha_0}^{\alpha_{\infty}} d\alpha' \langle (\vec{r}-\vec{r}_0)^2\rangle/2- M\ln\frac{\alpha_{\infty}\lambda^2}{2\pi}+M,
	\label{eq:svib}
\end{equation}}
where $\alpha_0$ and $\alpha_{\infty}$ respectively represent the limit for which the particles can escape from the reference configuration and the limit for which the quadratic potentials are infinitely steep. Again, for consistency, we set the de Broglie wavelength to $\lambda=1$.

\ft{In Fig.~\ref{figexampleMSDvsAlpha} we show the typical dependence on $\alpha$ of the integrand in equation \ref{eq:svib}. The method inherently has some arbitrariness in the choice of the lower limit of the integration $\alpha_0$: particles need to be able to move within their cages without escaping from them. From the shape of the curves in Fig. \ref{figexampleMSDvsAlpha} we identify a crossover regime at low $\alpha$ where the mean squared displacement rapidly increases as we decrease $\alpha$. Consistently with similar choices made in the literature \cite{angelani2007}, we choose a value of $\alpha_0$ close to the crossover region, $\alpha_0=100$.} 

Finally, the configurational entropy is then obtained as \begin{equation}
	\Sconf = \Sideal+\Sexc-\Svib : = M \sconf.
\end{equation}
\bibliographystyle{unsrt}
\bibliography{additions,allDropBox}

\begin{thebibliography}{10}

\bibitem{ito1999}
K.~Ito, C.~T. Moynihan, and C.~A. Angell.
\newblock Thermodynamic determination of fragility in liquids and a
  fragile-to-strong liquid transition in water.
\newblock {\em Nature}, 398:492--495, 1999.

\bibitem{debenedetti2001}
PG~Debenedetti and FH~Stillinger.
\newblock Supercooled liquids and the glass transition.
\newblock {\em Nature}, 410(6825):259--67, 2001.

\bibitem{royall2015physrep}
C.~P. Royall and S.~R. Williams.
\newblock The role of local structure in dynamical arrest.
\newblock {\em Phys. Rep.}, 560:1, 2015.

\bibitem{sciortino1999}
F.~Sciortino, W.~Kob, and P.~Tartaglia.
\newblock Inherent structure entropy of supercooled liquids.
\newblock {\em Phys. Rev. Lett.}, 83:3214--3217, 1999.

\bibitem{stillinger2002energy}
F.~H. Stillinger and P.~G. Debenedetti.
\newblock Energy landscape diversity and supercooled liquid properties.
\newblock {\em The Journal of chemical physics}, 116(8):3353--3361, 2002.

\bibitem{sciortino2005}
F.~Sciortino.
\newblock Potential energy landscape description of supercooled liquids and
  glasses.
\newblock {\em J. Stat. Mech.: Theory and Experiment}, page P05015, 2005.

\bibitem{turci2017nonequilibrium}
F.~Turci, C.~P. Royall, and T.~Speck.
\newblock Nonequilibrium phase transition in an atomistic glassformer: The
  connection to thermodynamics.
\newblock {\em Physical Review X}, 7(3):031028, 2017.

\bibitem{berthier2011}
L.~Berthier and G.~Biroli.
\newblock Theoretical perspective on the glass transition and amorphous
  materials.
\newblock {\em Rev. Mod. Phys.}, 83:587--645, 2011.

\bibitem{cammarota2012}
C.~Cammarota and G.~Biroli.
\newblock {Ideal glass transitions by random pinning.}
\newblock {\em Proc. Nat. Acad. Sci}, 109(23):8850--5, June 2012.

\bibitem{biroli2008}
G.~Biroli, J.~P. Bouchaud, A.~Cavagna, T.~S. Grigera, and P.~Verrochio.
\newblock Thermodynamic signature of growing amorphous order in glass-forming
  liquids.
\newblock {\em Nature Phys.}, 4:771--775, 2008.

\bibitem{berthier2012}
L.~Berthier and W.~Kob.
\newblock {Static point-to-set correlations in glass-forming liquids}.
\newblock {\em Phys. Rev. E}, 85:011102, 2012.

\bibitem{cammarota2013general}
C.~Cammarota.
\newblock A general approach to systems with randomly pinned particles:
  Unfolding and clarifying the random pinning glass transition.
\newblock {\em EPL (Europhysics Letters)}, 101(5):56001, 2013.

\bibitem{fullerton2014investigating}
C.~J Fullerton and R.~L Jack.
\newblock Investigating amorphous order in stable glasses by random pinning.
\newblock {\em Physical review letters}, 112(25):255701, 2014.

\bibitem{ozawa2015}
M.~Ozawa, W.~Kob, A.~Ikeda, and K.~Miyazaki.
\newblock Equilibrium phase diagram of a randomly pinned glass-former.
\newblock {\em Proc. Nat. Acad. Sci.}, 112:6914--6919, 2015.

\bibitem{gokhale2014}
S.~Gokhale, K.~H. Nagamanasa, R.~Ganapathy, and A.~K. Sood.
\newblock Growing dynamical facilitation on approaching the random pinning
  colloidal glass transition.
\newblock {\em Nature Comm.}, 5:4685, 2014.

\bibitem{gokhale2016}
S.~Gokhale, A.~K. Sood, and R.~Ganapathy.
\newblock Deconstructing the glass transition through critical experiments on
  colloids.
\newblock {\em Adv. Phys.}, 65(4):363---452, 2016.

\bibitem{gokhale2016jsm}
S.~Gokhale, K.~H. Nagamanasa, A.~K. Sood, and R.~Ganapathy.
\newblock Influence of an amorphous wall on the distribution of localized
  excitations in a colloidal glass-forming liquid.
\newblock {\em J. Stat. Mech.: Theory and Experiment}, page 074013, 2016.

\bibitem{Bianchi2010}
S.~Bianchi and Roberto Di~Leonardo.
\newblock Real-time optical micro-manipulation using optimized holograms
  generated on the gpu.
\newblock {\em Computer Physics Communications}, 181(8):1444Ð1448, 2010.

\bibitem{Bowman2011}
Richard~W. Bowman, V.~DÕAmbrosio, E.~Rubino, O.~Jedrkiewicz, P.~Di~Trapani, and
  Miles~J. Padgett.
\newblock Optimisation of a low cost slm for diffraction efficiency and ghost
  order suppression.
\newblock {\em European Physical Journal: Special Topics}, 199:149Ð158, 2011.

\bibitem{Thorneywork2014}
Alice~L. Thorneywork, Roland Roth, Dirk G. A.~L. Aarts, and Roel P.~A. Dullens.
\newblock Communication: Radial distribution functions in a two-dimensional
  binary colloidal hard sphere system.
\newblock {\em The Journal of Chemical Physics}, 140:161106, 2014.

\bibitem{Gray2015}
A.~T. Gray, E.~Mould, C.~P. Royall, and I.~Williams.
\newblock Structural characterisation of polycrystalline colloidal monolayers
  in the presence of aspherical impurities.
\newblock {\em Journal of Physics: Condensed Matter}, 27:194108, 2015.

\bibitem{Tamborini2015}
E.~Tamborini, C.~P. Royall, and P.~Cicuta.
\newblock Correlation between crystalline order and vitrification in colloidal
  monolayers.
\newblock {\em Journal of Physics: Condensed Matter}, 27(19):194124, 2015.

\bibitem{Cui2002}
B.~Cui, B.~Lin, S.~Sharma, and S.~A. Rice.
\newblock Equilibrium structure and effective pair interaction in a
  quasi-one-dimensional colloid liquid.
\newblock {\em The Journal of Chemical Physics}, 116:3119, 2002.

\bibitem{bernard2011}
E.~P. Bernard and W.~Krauth.
\newblock Two-step melting in two dimensions: First-order liquid-hexatic
  transition.
\newblock {\em Phys. Rev. Lett.}, 107:155704, Oct 2011.

\bibitem{assoud2010}
L.~Assoud, R~Messina, and H.~L\"{o}wen.
\newblock Ionic mixtures in two dimensions: From regular to empty crystals.
\newblock {\em EuroPhys. Lett.}, 89:36001, 2010.

\bibitem{crocker1995}
J.~C. Crocker and D.~G. Grier.
\newblock Methods of digital video microscopy for colloidal studies.
\newblock {\em J. Coll. Interf. Sci.}, 179:298--310, 1995.

\bibitem{jover2012p}
J~Jover, AJ~Haslam, A.~Galindo, G~Jackson, and E.~A. M{\"u}ller.
\newblock Pseudo hard-sphere potential for use in continuous molecular-dynamics
  simulation of spherical and chain molecules.
\newblock {\em The Journal of Chemical Physics}, 137(14):144505, 2012.

\bibitem{espinosa2013}
J.~R Espinosa, E.~Sanz, C.~Valeriani, and C.~Vega.
\newblock On fluid-solid direct coexistence simulations: The pseudo-hard sphere
  model.
\newblock {\em The Journal of chemical physics}, 139(14):144502, 2013.

\bibitem{ness2017effect}
C~Ness and A~Zaccone.
\newblock Effect of hydrodynamic interactions on the lifetime of colloidal
  bonds.
\newblock {\em Industrial \& Engineering Chemistry Research},
  56(13):3726--3732, 2017.

\bibitem{varga2016hydrodynamic}
Z~Varga and J~Swan.
\newblock Hydrodynamic interactions enhance gelation in dispersions of colloids
  with short-ranged attraction and long-ranged repulsion.
\newblock {\em Soft matter}, 12(36):7670--7681, 2016.

\bibitem{plimpton1995}
Steve Plimpton.
\newblock Fast parallel algorithms for short-range molecular dynamics.
\newblock {\em J. Comp. Phys.}, 117(1):1 -- 19, 1995.

\bibitem{Angelani2005}
Luca Angelani, Giuseppe Foffi, Francesco Sciortino, and Piero Tartaglia.
\newblock Diffusivity and configurational entropy maxima in short range
  attractive colloids.
\newblock {\em Journal of Physics: Condensed Matter}, 17(12):L113, 2005.

\bibitem{angelani2007}
L.~Angelani and G.~Foffi.
\newblock Configurational entropy of hard spheres.
\newblock {\em J. Phys.: Condens. Matter}, 17:256207, 2007.

\bibitem{frenkel}
D.~Frenkel and B.~Smit.
\newblock {\em Understanding molecular simulation: from algorithms to
  applications}.
\newblock Elsevier, 2nd edition, 2002.

\bibitem{piaggi2017}
P.~M Piaggi, O.~Valsson, and M.~Parrinello.
\newblock Enhancing entropy and enthalpy fluctuations to drive crystallization
  in atomistic simulations.
\newblock {\em Phys. Rev. Lett.}, 119(1):015701, 2017.

\bibitem{illing2017mermin}
B~Illing, S~Fritschi, H~Kaiser, C~L Klix, G~Maret, and P~Keim.
\newblock Mermin--wagner fluctuations in 2d amorphous solids.
\newblock {\em Proceedings of the National Academy of Sciences},
  114(8):1856--1861, 2017.

\bibitem{nagamanasa2015}
S.~Nagamanasa, K.~H.Nagamanasa, A.~K. Sood, and R.~Ganapathy.
\newblock Direct measurements of growing amorphous order and non-monotonic
  dynamic correlations in a colloidal glass-former.
\newblock {\em Nature Phys}, 11:403--408, 2015.

\bibitem{pastore2015connecting}
R~Pastore, M~P Ciamarra, G~Pesce, and A~Sasso.
\newblock Connecting short and long time dynamics in hard-sphere-like colloidal
  glasses.
\newblock {\em Soft Matter}, 11(3):622--626, 2015.

\bibitem{starr2002}
F~W Starr, S~Sastry, J~F Douglas, and S~C Glotzer.
\newblock What do we learn from the local geometry of glass-forming liquids?
\newblock {\em Physical review letters}, 89(12):125501, 2002.

\bibitem{hocky2014}
G.~M. Hocky, D.~Coslovich, A.~Ikeda, and D.~Reichman.
\newblock Correlation of local order with particle mobility in supercooled
  liquids is highly system dependent.
\newblock {\em Phys. Rev. Lett.}, 113:157801, 2014.

\bibitem{pastore2017}
R~Pastore, G~Pesce, A~Sasso, and M~Pica~Ciamarra.
\newblock Cage size and jump precursors in glass-forming liquids: Experiment
  and simulations.
\newblock {\em The Journal of Physical Chemistry Letters}, 8(7):1562--1568,
  2017.

\bibitem{eckmann2008}
J.-P. Eckmann and I.~Procaccia.
\newblock Ergodicity and slowing down in glass-forming systems with soft
  potentials: No finite-temperature singularities.
\newblock {\em Phys. Rev. E}, 78:011503, 2008.

\bibitem{cammarota2011}
C.~Cammarota and G.~Biroli.
\newblock Ideal glass transitions by random pinning.
\newblock {\em ArXiV:cond-mat}, page 1106.5513v1, 2011.

\bibitem{sausset2011}
F.~Sausset and D.~Levine.
\newblock Characterizing order in amorphous systems.
\newblock {\em Phys. Rev. Lett.}, 107:045501, 2011.

\bibitem{russo2015}
J.~Russo and H.~Tanaka.
\newblock Assessing the role of static length scales behind glassy dynamics in
  polydisperse hard disks.
\newblock {\em Proc. Nat. Acad. Sci.}, 112:6920---6924, 2015.

\bibitem{banerjee2016effect}
A.~Banerjee, M.~K. Nandi, S.~Sastry, and S.~M. Bhattacharyya.
\newblock Effect of total and pair configurational entropy in determining
  dynamics of supercooled liquids over a range of densities.
\newblock {\em The Journal of chemical physics}, 145(3):034502, 2016.

\bibitem{berthier2017}
Ludovic Berthier, Patrick Charbonneau, Daniele Coslovich, Andrea Ninarello,
  Misaki Ozawa, and Sho Yaida.
\newblock Configurational entropy measurements in extremely supercooled liquids
  that break the glass ceiling.
\newblock {\em Proceedings of the National Academy of Sciences}, page
  201706860, 2017.

\end{thebibliography}
\end{document}